\documentstyle[preprint,tighten,12pt,aps,epsf]{revtex}
\newcommand{\dss}{D^{\ast \ast}}

\newcommand{\at}{a_{2}(1320)}
\newcommand{\aoo}{a_{1}(1260)}
\newcommand{\az}{a_{0}(1450)}
\newcommand{\bo}{b_{1}(1235)}
\newcommand{\ft}{f_{2}(1270)}
\newcommand{\fo}{f_{1}(1285)}
\newcommand{\fz}{f_{0}(1370)}
\newcommand{\ho}{h_{1}(1170)}
\newcommand{\kt}{K_{2}^{\ast}(1430)}
\newcommand{\kl}{K_{1}(1270)}
\newcommand{\kz}{K_{0}^{\ast}(1430)}
\newcommand{\kh}{K_{1}(1400)}
\newcommand{\fts}{f_{2}^{'}(1525)}
\newcommand{\fos}{f_{1}(1510)}
\newcommand{\fzs}{f_{0}(1370)}
\newcommand{\hos}{h_{1}(1380)}
\newcommand{\kq}{K (1460)}
\newcommand{\ksq}{K^{\ast} (1410)}
\newcommand{\pq}{\pi (1300)}
\newcommand{\euq}{\eta_{u} (1295)}
\newcommand{\esq}{\eta_{s} (1490)}
\newcommand{\rqb}{\rho (1450)}
\newcommand{\oq}{\omega (1420)}
\newcommand{\btab}{\begin{tabbing}}
\newcommand{\etab}{\end{tabbing}}

\newcommand{\gapproxeq}{\lower.7ex\hbox{$\;\stackrel{\textstyle>}{\sim}\;$}}
\newcommand{\lapproxeq}{\lower.7ex\hbox{$\;\stackrel{\textstyle<}{\sim}\;$}}

\begin{document}

\title{ Hybrid Meson Decay Phenomenology }

\author{Philip R. Page$^1$, Eric S. Swanson$^{2,3}$, and Adam P. Szczepaniak$^4$}

\address{$^1$\
  Theoretical Division,
  Los Alamos National Laboratory,
  Los Alamos, NM 87545\\
         $^2$
  Department of Physics,
  North Carolina State University,
  Raleigh, North Carolina 27695-8202\\
         $^3$\
  Jefferson Laboratory,
  12000 Jefferson Avenue,
  Newport News, VA 23606\\
      $^4$\
  Department of Physics,
  Indiana University,
  Bloomington, Indiana   47405-4202\\
        }

\maketitle

\begin{abstract}
The phenomenology of a newly developed model of hybrid meson decay is developed. The decay
mechanism is based on the heavy quark expansion of QCD and the strong coupling flux tube
picture of nonperturbative glue. A comprehensive list of partial decay widths of a wide
variety of light, $s\bar s$, $c\bar c$, and $b \bar b$ hybrid mesons is presented.
Results which appear approximately universal are highlighted along with those which 
distinguish different hybrid decay models. Finally, we examine several interesting hybrid
candidates in detail.

\end{abstract}


\section{Introduction}

Quantum Chromodynamics at low energy remains enigmatic chiefly because of an almost complete
lack of knowledge of the properties of soft glue. Glue must certainly be understood
if phenomena such as color confinement, mass generation, and dynamical symmetry breaking are 
to be understood. The discovery and explication of hadrons with excited gluonic degrees of 
freedom is clearly an important step in this process.
Furthermore, the search for nonperturbative glue, in particular as manifested in hybrid 
mesons, would be greatly
facilitated by a rudimentary knowledge of the hybrid spectrum and decay
characteristics. 
Although it appears that lattice estimates of light quenched
hybrid masses are forthcoming\cite{lacock96}, hadronic decays remain difficult
to calculate on the lattice. Thus one is forced to rely on model estimates
of the couplings of hybrids to ordinary mesons.

Historically, there have been two 
approaches to such estimates. The first assumes that hybrids are 
predominantly quark-antiquark states with an additional constituent 
gluon\cite{HM} and that decays proceed via constituent gluon 
dissociation\cite{Orsay}. The second assumes that hybrids are 
quark-antiquark states moving on an adiabatic surface generated by an excited 
``flux tube" configuration of glue\cite{IP}. Decays then proceed by a 
phenomenological pair production mechanism (the ``$^3P_0$ model") coupled 
with a flux tube overlap\cite{IKP}. An important feature of this model is that 
the quark pair creation vertex is uncorrelated with the gluonic modes of the
hybrid.

A third possibility for hybrid decay has been recently introduced
\cite{ss3}. This model also assumes flux tube hybrids but employs a different
decay vertex. The vertex is constructed by using the heavy quark
expansion of the Coulomb gauge QCD Hamiltonian to identify relevant operators. 
The gluonic portion of these are then evaluated using a slightly
extended version of the flux tube model of Isgur and Paton\cite{IP}.
The essential new feature is that the gluon field
operator is expressed in terms of the nonperturbative phonon modes
of the flux tube model rather than traditional plane waves.

This paper begins with a review of the development of the decay model of Ref. \cite{ss3}
and describes in detail several issues which arise in converting the amplitudes to decay
widths. We then summarize the main general features of the model and compare these with
the flux tube decay model of Isgur, Kokoski, and Paton (hereafter referred to as IKP). 
The main portion of this work is
a comprehensive review of the decay modes of all low lying isovector, isoscalar, $s \bar s$, $c 
\bar c$, and $b \bar b$ $2^{\mp \pm}$, $1^{\pm\pm}$, $1^{\mp\pm}$, and $0^{\mp\pm}$ hybrids.
A detailed discussion of interesting features in the phenomenology of these states follows.

\section{Hybrid Decay Amplitude}

The first step in the construction of any hybrid decay model is determining what is meant by a 
hybrid. 
We stress that choosing a model of hybrids with the correct degrees of freedom is crucial because
decays probe the internal structure of the participating particles. Thus for example, in the flux tube
model low lying vector hybrids must have the quarks in a spin singlet and this implies that 
vector hybrids may not decay to a pair of spin zero mesons (see below for further discussion of
this point). However, this need not be true in
a model which assigns hybrid quantum numbers differently (for example, it is possible to 
construct spin one vector hybrids in constituent glue models). 
In this work, we choose to employ a slightly modified version of the flux tube model hybrids
of Isgur and Paton, as described in Refs. \cite{ss3,ss2}. Recent lattice calculations of adiabatic hybrid potential surfaces show that the
flux tube model does a good job of describing the level orderings and degeneracies apparent in the
data (although it does not reproduce many details)\cite{JKM}. Thus one may be confident that the
model captures the essential features of (heavy) hybrid structure necessary for the construction
of a viable decay model.

The flux tube model of Isgur and Paton \cite{IP} 
is extracted from the strong coupling limit of the QCD lattice 
Hamiltonian. The Hamiltonian is first split into blocks of distinct
``topologies" (in reference to possible gauge invariant flux tube 
configurations)  and then adiabatic and  small oscillation approximations
of the flux tube dynamics are made to arrive at an N-body discrete string-like 
model Hamiltonian
for gluonic degrees of freedom. This is meant to be operative at intermediate 
scales $a \sim b^{-1/2}$ where the strong coupling is of order unity.
The lattice spacing is denoted by $a$, the string tension by $b$, and there are $N$ 
``beads'' (or links) evenly spaced between the $Q \bar Q$ pair. Diagonalizing 
the flux tube Hamiltonian yields phonons, $\alpha^a_{m,\lambda}$, which 
are labelled by their color ($a$), mode number ($m$), and polarization ($\lambda$).
A hybrid may be built
of $n_{m\lambda}$ phonons in the m'th mode with polarization $\lambda = \pm$.
In particular, hybrid states with a single phonon excitation are constructed as

\begin{equation}
\vert H \rangle \sim   \int d {\bf r} 
\, \varphi_H(r) \, \chi^{PC}_{\Lambda,\Lambda'}\, 
D^{L_H*}_{M_L,\Lambda}(\phi,\theta,-\phi) \, T^a_{ij}\, b_i^\dagger({\bf r}/2) 
d_j^\dagger(-{\bf r}/2) \alpha^{a\dagger}_{m,\Lambda'} \vert 0 \rangle.
\end{equation}

\noindent
Spin and flavor indices have been suppressed and color indices are explicit.
The factor $\chi^{PC}_{\Lambda,\Lambda'}$ in the hybrid wavefunction projects
onto states of good parity and 
charge conjugation. The quantum numbers of these 
states are given by
$P = \eta_{PC} (-)^{L_H + 1}$ and $C = \eta_{PC} (-)^{L_H + S_H + N}$ where
$\eta_{PC} = \chi^{PC}_{-1,-1} = \pm 1$ and 
$N = \sum_m m(n_{m+} + n_{m-})$. 
These expressions differ from Isgur and 
Paton\cite{IP} because we have adopted the standard definitions for the polarization
vectors and the Wigner
rotation matrix, following the Jacob-Wick conventions.
We shall consider low-lying hybrids only so that $m=1$ in what follows.

It remains to specify the structure of the decay operator. 
To leading order in the hopping parameter and strong coupling
expansion,  
one can show that the operator for producing a $q({\bf r}_q)\bar 
q({\bf r}_{\bar q})$ pair has the following structure \cite{grom},

\begin{equation}
 F_{q\bar q} \propto 
 e^{-m |{\bf r}_{q\bar q}|}  b^{\dagger}({\bf r}_q)
{\bf r}_{q\bar q} \cdot \bbox{\sigma} 
d^{\dagger}({\bf r}_{\bar q}).
\end{equation}
The dependence on the relative distance, ${\bf r}_{q\bar q} = 
{\bf r}_q - {\bf r}_{\bar q}$, comes from integrating  $n=  |{\bf
  r}_{q\bar q}|/a$ products of the link operators from the kinetic
term 
\begin{equation}
 K = -\kappa m \sum_{n,\mu} \bar
  \psi_{n}(1+\gamma^\mu)U_{n,\mu}\psi_{n+\mu} + H.c,
\end{equation} 
  over a straight line in direction of ${\bf r}_{q\bar q}$. 
The prefactor $ e^{-m |{\bf r}_{q\bar q}|} = (2 \kappa)^n$ can be
 identified with the Schwinger tunneling factor for pair production in
 an external field of the parent $q\bar q$ meson. 
 In our picture, hybrids are characterized by excitations of the
 gluonic field. We will therefore assume that hybrid decays can proceed
 through local deexcitation of this field rather then by quark
 tunneling in the external field of the meson source. Thus  
 the expectation value of the gluon operator in $K$ between excited 
 (hybrid) and
 deexcited (low lying meson decay products) is used to obtain the effective
 $q\bar q$ production operator. 
In \cite{ss3} the chromoelectric $\bf E$ and ${\bf B}$ fields have been
mapped onto the flux tube space of gluon excitations described by the
phonon operators. Using these expressions together with ${\bf
  E}=-d{\bf A}/dt$ one then obtains 
\begin{equation}
A^a_\lambda({\bf x}_n,t) = {-i \over a\sqrt{(N+1)}} \sum_m \cos({m \pi \over
N+1}n) {1 \over \sqrt{a\omega_m}} \left(\alpha^a_{m \lambda} {\rm e}^{-i \omega_m t}
 - \alpha^{a \dagger}_{m \lambda} {\rm e}^{i \omega_m t} \right)
\end{equation}

This yields the following effective decay operator

\begin{equation}
H_{int} = {i g a^2 \over \sqrt{\pi}}  \sum_{m,\lambda} \int_0^1 d \xi 
\cos(\pi \xi)  T^a_{ij}\, h^\dagger_i(\xi {\bf r}_{Q\bar Q}) 
{\bbox\sigma}\cdot \hat{\bf e}_{\lambda}(\hat{\bf r}_{Q\bar Q}) \left(\alpha^a_{m \lambda} 
 - \alpha^{a \dagger}_{m \lambda} \right) \chi_j(\xi {\bf r}_{Q\bar Q}),
\end{equation}

\noindent 
where the $\hat{\bf e}(\hat{\bf r})$ are polarization vectors orthogonal to $\hat {\bf r}$.
The integral is defined along the $Q\bar Q$
axis only. Integration over the transverse directions yields the 
factor $a^2$ which may be interpreted as the transverse size of the flux tube. 
Note that the phonon operators represent gluonic excitations which are 
perpendicular to the $Q\bar Q$
axis. Although this appears problematical in traditional perturbation theory,
it is required here because, in the adiabatic limit, the gluonic field 
configuration must be defined in terms of the quark 
configuration and therefore the field expansion of the vector potential
depends on the quark state under consideration. 

The decay amplitude for a hybrid $H$ into mesons $A$ and $B$ is then given 
by:

\begin{eqnarray}
\langle H \vert H_{int}\vert AB\rangle &=& i {g a^2\over \sqrt{\pi}} {2\over 3}
  \int_0^1 d \xi \int d{\bf r} \cos(\pi \xi) \sqrt{2L_H+1\over 4 \pi} 
  {\rm e}^{i {\bf p}\cdot {\bf r}\over 2} \varphi_H(r) \varphi^*_A(\xi{\bf r}) 
  \varphi^*_B((1-\xi){\bf r}) \cdot \nonumber \\
  && \left[ {\cal D}^{L_H*}_{M_L \Lambda}(\phi,\theta,
  -\phi) \chi_{\Lambda,\lambda}^{PC} \hat {\bf e}_{\lambda}(\hat {\bf r}) \cdot
  \langle \bbox{\sigma} \rangle \right]
\end{eqnarray}

\noindent
where $\langle \bbox{\sigma}\rangle$ is the matrix element of the Pauli matrices 
between quark spin wavefunctions,
\begin{equation}
\langle \bbox{\sigma} \rangle = \langle {1\over 2} s {1\over 2} \bar s\vert
S_H M_H\rangle \, \langle {1\over 2} s {1\over 2} \bar s_A\vert S_A M_A\rangle
\, \langle {1\over 2} s_B {1\over 2} \bar s\vert S_B M_B\rangle
 {\bbox \sigma}_{S_B \bar S_A}.
\end{equation}
This  amplitude should be multiplied by
the appropriate flavor overlap and symmetry factor.

The evaluation of the matrix elements is greatly facilitated by performing the angular
integrals analytically. This may be achieved 
through use of the relations $\hat {\bf e}_\Lambda(\hat r) = \sum_\lambda D^1_{\lambda \Lambda}(\phi,
\theta,-\phi) \hat {\bf e}_\lambda(\hat z)$ and

\begin{equation}
\hat {\bf e}_\lambda(\hat z) \bbox{\sigma}_{S_B \bar S_A}  = -\sqrt{2} \langle {1\over 2} 
S_B {1\over 2} S_A \vert 1 \lambda \rangle.
\end{equation}

\noindent
The resulting expression completely factorizes from the radial and flux tube integrals except for
a trivial dependence on the wave in the final channel, greatly simplifying the algebra.

We note the following general properties of the decay amplitude. The operator 
is nonzero only along the hybrid $Q\bar Q$ axis -- as follows from the structure
of the interaction Hamiltonian. Thus $q\bar q$ creation occurs on a line joining
the original $Q\bar Q$ quarks, smeared over the transverse size of the flux
tube. This is in contrast to the model of IKP which has transverse extent and a node along the 
$Q\bar Q$
axis. Furthermore the spin operator contracts with the flux tube
phonon polarization vector, which is absent in the IKP model.
Finally, the decay amplitude vanishes when the final mesons are identical
due to the nodal structure in the vector potential. 
This is true for any single-phonon hybrid in an odd mode.
Thus one obtains the selection rule:
low-lying hybrids do not decay to identical mesons. This subsumes the 
selection rule of IKP so that none of their qualitative conclusions are 
changed. However we also predict, for example, that hybrids do not decay
to pairs of identical P-wave mesons. 
This rule has recently been shown to be more general
than specific models \cite{page97sel2}. The preferred decay channels are to $S + P$ --wave 
pairs\cite{Orsay2,IKP}. 
We stress that the selection rule forbidding $S + S$ --wave final states
no longer operates if the internal structure or size of the two S--wave states
differ \cite{ss3,page95hybrid}. 

Another rule, the ``spin selection" rule, exists: if the $q\bar{q}$ in either hybrid or 
conventional mesons
are in a net spin singlet configuration then decay into final states consisting only of spin singlet
states is forbidden. This rule follows because pair creation is spin-triplet.
It appears to be a universal feature in all
non--relativistic decay models.

For $J^{PC}=1^{--}$ states this selection rule
distinguishes between conventional vector mesons
which are $^3S_1$ or $^3D_1$ states and hybrid vector mesons where the
$q\bar{q}$ are coupled to a spin singlet. For example, it implies that in the decay of
hybrid $\rho_H$, the channel
$\pi h_1$ is forbidden whereas $\pi a_1$ is allowed; this
is quite opposite to the case of $^3L_1$ conventional mesons where the
$\pi a_1$ channel is relatively suppressed and $\pi h_1$
is allowed \cite{kokoski87,page96rad}. The extensive analysis of data
in Ref. \cite{donnachie94} revealed the clear presence of $\rho(1450)$ \cite{pdg96}
 with a strong $\pi a_1$ mode but no sign of $\pi h_1$,
in accord with the hybrid situation. 
 
There are a number of amplitudes
that vanish for the SHO wave functions employed here in addition to those governed by
the selection rules above. Some of these decays vanish simply  
due to quantum numbers, e.g.
$J^{PC} = 0^{-+}$ to two vector mesons (see the proof in Appendix 1 of Ref. \cite{page97pho}). 

Some amplitudes vanish in both this work and the IKP model. 
These include all F--wave amplitudes for hybrid decay to two S--wave mesons, and all
G--wave amplitudes. Also, $0^{-+}, 1^{+-}$ hybrid decays to two  vector mesons
vanishes.

In addition, the decays $2^{-+}$ and $1^{+-}\rightarrow
1^{+-}\; 0^{-+},\;$; 
$1^{++}\rightarrow 0^{++}\; 0^{-+}$; and $ 0^{+-}\rightarrow 1^{++}\; 0^{-+}$
vanish. Alternatively, in the IKP model $2^{-+}\rightarrow 1^{++}\; 0^{-+}$ and
$1^{-+}\rightarrow 2^{++}\; 0^{-+}$ vanish.

\section{Hybrid Meson Widths}

The final step is to calculate hybrid widths. This involves choosing
prescriptions for evaluating the decay phase space, the vertex coupling $g a^2$, 
and wavefunction parameters.

The choice of the  appropriate phase space 
is, unfortunately, a difficult issue to resolve
(it is discussed extensively in \cite{geiger94}).
For example, in our conventions standard relativistic phase space evaluates to

\begin{equation}
  {\rm (ps)} = 2\pi k {E_A E_B \over m_H} 
\end{equation}

\noindent
where $E_A$ is the energy of meson $A$ in the final state. This can differ substantially
>from the nonrelativistic version:

\begin{equation}
  {\rm (ps)} = 2\pi k {m_A m_B \over (m_A + m_B)} 
\end{equation}

\noindent
especially when pions are in the final state. Finally, we mention a third possibility
employed by Kokoski and Isgur \cite{kokoski87}, called the `mock meson' method. The
authors use

\begin{equation}
  {\rm (ps)} = 2\pi k {M_A M_B \over M_H} 
\end{equation}

\noindent
where $M_A$ refers to the `mock meson' mass of a state. This is defined to be the
hyperfine-splitting averaged meson mass. In practice, the numerical result is little
different from the relativistic phase space except for the case of the pion, where a
mock mass of $M_\pi = 0.77$ GeV is used. The net effect on low lying meson decays is to 
enhance the decay for processes with pions in the final state by a factor of 
$M_\pi/E_\pi$ for each pion in the final state. This procedure improved the fit to
experimental data substantially. In fact, it is generally true that the $^3P_0$ model
(with relativistic phase space)
fits the data quite well except for the case where pions are in the final state.

We have adopted a different approach to phase space which also solves this problem 
and which we believe is better physically motivated.  We suggest \cite{swan98}
that the root of the problem lies in the Goldstone boson nature of the pion. This 
implies that a pion is not a simple $Q \bar Q$ state, but rather is collective in
nature. An explicit way to incorporate this physics into a constituent quark model
has been suggested by several groups\cite{Orsay3,ss2}.  The method relies on 
constructing a nontrivial vacuum for QCD which breaks chiral symmetry. The pion may
then be manifested as a Goldstone mode by using the random phase approximation (RPA)
to construct it. The point of interest to the current discussion is that in the
random phase approximation the pion wavefunction contains backward moving pieces.
These pieces allow new contributions to meson 
decay diagrams when pions are in the final state. In the chiral limit, the net
result is quite simple: amplitudes with two pions in the final state should be
multiplied by 3 (over the naive quark model result), while those with a single pion 
in the final state should be multiplied by 2.  The efficacy of this prescription 
is illustrated in Table 1. 

\begin{table}
\caption{$^3P_0$ couplings needed to reproduce experimental widths.}
\begin{tabular}{ccccc}
  & $\rho \rightarrow \pi\pi$ & $b_1 \rightarrow \omega \pi$ & $a_1 \rightarrow \rho\pi$ &
$\pi_2 \rightarrow \rho\pi$ \cr
no RPA & 0.71 & 0.53 & 0.46 & 0.42 \cr
RPA & 0.24 & 0.26 & 0.23 & 0.21 \cr
\end{tabular}
\label{tableI}
\end{table}

\noindent
As can be seen, the improvement is dramatic.
Precisely the same argument applies to hybrid decays. Thus our prescription is as 
follows: use relativistic phase space and the RPA pion factors mentioned above to
arrive at the final decay amplitudes.

The work of IKP was greatly expanded in Close and Page\cite{page95hybrid}; since one of
the purposes of this work is to compare this model with IKP, we have both quoted
the results of Close and Page below and have used their meson and hybrid meson 
wavefunction parameters as our ``standard parameters" (these are discussed in the Appendix).
Note that in order to calculate the IKP model predictions given below, we use the same
normalization as in Ref. \protect\cite{page95hybrid}, which corresponds to
the $^3P_0$ pair creation parameter $\gamma_0 = 0.39$ favored
for mock meson phase space \protect\cite{kokoski87,geiger94}. Although
$\gamma_0 = 0.53$ is preferred for relativistic phase space
\protect\cite{geiger94}, Ref. \protect\cite{page96rad} used $\gamma_0 = 0.4$ for 
high mass meson resonances. We simply choose to retain $\gamma_0 = 0.39$.

The normalization of this model is fixed 
to give the same average width as the IKP model for the
decays of isovector hybrids to $\eta\pi,\;\eta^{'}\pi$ and $\rho\pi$ 
with the ``standard parameters''. This yields a coupling of $ga^2 = 1.78$ GeV$^{-2}$.
These particular decay modes were chosen because
the two models can analytically be shown to mimic the
predictions of each other in decays to two ground state or 
radially excited S--wave final states. Thus decays to
these final states may be regarded as ``model invariant''.
Finally, as discussed above, we note that the absolute widths in the IKP model
could be up to $(0.53/0.39)^2 \approx 2$ times bigger than the widths
quoted here. Furthermore, since phase
space conventions and absolute magnitude conventions have changed
since former IKP model calculations \cite{page95hybrid} care should be
taken with comparisons. Indeed, the authors of IKP state that a model error of  
(an additional) factor of 2 should be allowed for in their predicted widths.

To make contact with the original development of this model\cite{ss3} and to
illustrate the parameter dependence of the model predictions, we also employ the 
parameters of Ref. \cite{ss3} as an ``alternative parameter" set.  
This set was normalized
to the experimental decay pattern of the hybrid meson candidate $\pi(1800)$, yielding
\footnote{Note that Ref. \cite{ss3} did not use the RPA
pion prescription. The value of the coupling quoted here corrects this. None of the results
of that paper change.}
$ga^2 = 1.28$ GeV$^{-2}$. These parameters are also listed in the Appendix.

Simple harmonic oscillator (SHO) wavefunctions are used throughout for the final state
mesons. This is typical of decay calculations and it has been demonstrated that using
Coulomb + linear wavefunctions does not change the results significantly\cite{geiger94,kokoski87}.
We have taken the following masses for the  $u\bar{u}$, $s\bar{s}$,$c\bar{c}$, and $b\bar{b}$
hybrids: 1.8, 2.0, 4.1, and  10.7 GeV respectively. Masses
for known mesons are taken from Ref. \cite{pdg96} and otherwise
>from Ref. \cite{isgur85}.
The quark model assignments for the mesons are those of the
PDG tables \protect\cite{pdg96}. The $\fz$ is assumed to be the scalar 
$\frac{1}{\sqrt{2}}(u\bar{u}+d\bar{d})$
state. We assume the $J^{PC} =
2^{++},1^{++},0^{++},1^{+-}$ 
$s\bar{s}$ mesons to be $\fts , \fos , \fzs , \hos $ respectively.
Thus  $\fzs$ denotes a generic scalar state at 1.37 GeV,
containing either light quarks or $s\bar{s}$, depending
on the context. 

The flavor structure of the $\eta$ is taken to be
$\protect\sqrt{\frac{1}{2}}(\protect\sqrt{\frac{1}{2}}(u\bar{u}+d\bar{d})-s\bar{s})$
at 547 MeV and $\eta^{'}$ is 
$\protect\sqrt{\frac{1}{2}}(\protect\sqrt{\frac{1}{2}}(u\bar{u}+d\bar{d})+s\bar{s})$ at 958 MeV.
The $\eta_{u} (1295)$ and $\eta_{s} (1490)$ are assumed to be $\protect\sqrt{\frac{1}{2}}(u\bar{u}+d\bar{d})$
and $s\bar{s}$ respectively, with $\eta_{s} (1490)$ the second $\eta (1440)$ peak at
1490 MeV. $\kq$ is not well--established. $\dss(2^+)$ denotes
the PDG state $D^{\ast}_2(2460)$. $\dss({1^{+}_L})$ and
$\dss({1^{+}_H})$ are the low and high mass $1^+$ states respectively.
The high mass state can be identified with the PDG state $D_1(2420)$.

As stated earlier, we employ  relativistic phase space and RPA pion factors
and work in the narrow resonance approximation. We also extend the RPA prescription
to kaons and $\eta$s; but not to the $\eta'$. 
Decay modes include all possible charge combinations, e.g. $\rho\pi$
means $\rho^+\pi^-,\;\rho^0\pi^0$ and $\rho^-\pi^+$. 

In the following tables we present the 
dominant widths for hybrid $H \rightarrow A B$ for various
$J^{PC}$ hybrids in partial wave $L$. 
Column 1 indicates the $J^{PC}$ of the
hybrid, column 2 the decay mode and column 3, $L$. 
In columns 4, 5, 6 and 8 we indicate predictions of this model. Column
6 uses the ``standard parameters'' used throughout the text and
defined in the Appendix. Column 5 uses the same parameters, except that all
hybrids are assumed to be $0.2$ GeV heavier (and the $c\bar{c}$ hybrids
$0.3$ GeV heavier to put them above the $D^{**}D$ thresholds
at approximately $4.3$ GeV). Column 4 uses
the ``alternative parameters''. Columns
4 and 6 should hence be compared to estimate parameter sensitivity of
our predictions. For hybrid decays to two ground state S--wave mesons we indicate the
``reduced width'' in column 8. This is the width divided by
the dimensionless ratio
$(\beta_A^2-\beta_B^2)^2/(\beta_A^2+\beta_B^2)^2$,
where $\beta$ is the inverse radius of the SHO wave function \protect\cite{page95hybrid}.
It gives a measure of how strong the decay is with the difference
of the wave functions explicitly removed. In column 7 we give IKP
model predictions for the ``standard parameters'', so that columns 6
and 7 should be compared when this model is compared with the IKP
model.

As stated earlier, we omit F--wave amplitudes for hybrid to two S--wave mesons, and all
G--wave amplitudes, since these vanish in both models.
We do not list decays with two S--wave mesons in the
final state which have identical wave functions (e.g. $\pi\pi,\;
\rho\rho$), since these amplitudes vanish due to the ``S+S'' selection rule.
The symbol ``\O'' indicates that an amplitude is exactly zero,
not only numerically small. Finally, a dash indicates that a decay mode
is below threshold (recall that we work in the narrow width approximation).

\begin{table}
\caption{Isovector Hybrid Decay Modes}
\begin{tabular}{lccccccc}
         &                &   &  alt & high mass & standard & IKP & reduced \\
\hline
$2^{-+}$ & $\rho\pi  $     & P &    9 &  16  &  13  &  12  &  57  \\
         & $K^{\ast} K$    & P &    1 &   5  &   2  &   1  &  17  \\
         & $\rho\omega$    & P &    0 &   0  &   0  &   0  &  20  \\
         & $ \ft\pi  $     & S &   19 &  10  &  9   &  14  &      \\
         &                 & D &   .1 &  .2  & .05  &  11  &      \\
         & $ \fo\pi  $     & D &   .1 &  .3  & .06  &  \O  &      \\
         & $ \fz\pi  $     & D &   .02&  .08 & .01  &  .6  &      \\
         & $ \bo\pi  $     & D & \O   & \O   & \O   &  20  &      \\
         & $ \at\eta  $    & S &  --  &  7   &  --  &  --  &      \\
         &                 & D &  --  & .01  &  --  &  --  &      \\
         & $ \aoo\eta  $   & D &   0  & .05  &   0  &   0  &      \\
         & $ \az\eta  $    & D & --   &  0   &  --  &  --  &      \\
         & $ \kt K $       & S & --   & 11   & --   &  --  &      \\
         &                 & D & --   &  0   & --   &  --  &      \\
         & $ \kl K $       & D & 0    & .01  &  0   & .02  &      \\
         & $ \kz K $       & D & --   & 0    & --   &  --  &      \\
         & $ \kh K $       & D & --   & 0    & --   &  --  &      \\
         & $ \rqb \pi $    & P &  .8  &  12  &   3  &   2  &      \\
         & $ \ksq K   $    & P &   -- &   1  &  --  &  --  &      \\
         & $ \Gamma$       &  &  30   &  63  &  27  & 59   &      \\
\tableline
$1^{-+}$ & $ \eta \pi    $ & P &    0 &   .02&   .02& .02  &  99  \\
         & $ \eta^{'} \pi$ & P &    0 &   .01&   .01& 0    &  30  \\
         & $ \rho \pi    $ & P &    9 &  16  &  13  &  12  &  57  \\
         & $ K^{\ast} K  $ & P &    1 &   5  &   2  &   1  &  17  \\
         & $ \rho \omega $ & P &    0 &   0  &   0  &   0  &  13  \\
         & $ \ft\pi  $     & D & .2   &  .5  &  .1  & \O   &      \\
         & $ \fo\pi  $     & S &  18  &  10  &  9   &  14  &      \\
         &                 & D &   .06&  .2  & .04  &   7  &      \\
         & $ \bo\pi      $ & S &  78  &  40  &  37  &  51  &      \\
         &                 & D &   2  &   3  &   1  &  11  &      \\
         & $ \at\eta  $    & D & --   & .02  &  --  &  --  &      \\
         & $ \aoo\eta  $    & S & 5    &   7  &   3  &   8  &      \\
         &                 & D & 0    & .01  &  0   &  .01 &      \\
         & $ \kt K $       & D & --   & 0    & --   &  --  &      \\
         & $ \kl K $       & S &  4   & 7    &  2   &   6  &      \\
         &                 & D & 0    & .2   & 0    & .04  &      \\
         & $ \kh K $       & S & --   & 33   & --   &  --  &      \\
         &                 & D & --   &  0   & --   &  --  &      \\
         & $ \pq \eta $    & P & --   &   5  &  --  &  --  &      \\
         & $ \euq \pi $    & P & 3    &  27  &  11  &   8  &      \\
         & $ \kq K    $    & P & --   &  .8  &  --  &  --  &      \\
         & $ \rqb \pi $    & P &  .8  &   12 &    3 &   2  &      \\
         & $ \ksq K   $    & P &  --  &   1  &  --  &  --  &      \\
         & $\Gamma $       &   & 121  & 168 & 81    &  117 &      \\
\hline
$1^{--}$ & $ \omega \pi  $ & P &   9  &  16  &  13  &  12  &  57  \\
         & $\rho\eta     $ & P &   4  &   9  &   6  &   4  &  30  \\
         & $\rho\eta^{'} $ & P &  .1  &   1  &   .2 &  .1  &   1  \\
         & $ K^{\ast} K  $ & P &   3  &   9  &   5  &  3   &  34  \\
         & $ \at\pi  $     & D &  .5  &   2  &   .3 &  16  &      \\
         & $ \aoo\pi  $     & S &   78 &  41  &  37  &  51  &      \\
         &                 & D &   .4 &   .8 &   .2 &  11  &      \\
         & $ \ho\pi  $     & S &      &      &   \O &      &      \\
         &                 & D &      &      &  \O  &      &      \\
         & $ \bo\eta     $ & S &      &      &  \O  &      &      \\
         &                 & D &      &      &  \O  &      &      \\
         & $ \kt K $       & D & --   & 0    & --   &  --  &      \\
         & $ \kl K $       & S &  6   &  12  &  4   &  11  &      \\
         &                 & D &  0   & .01  & 0    &  0   &      \\
         & $ \kh K $       & S & --   &  17  & --   &  --  &      \\
         &                 & D & --   & 0    & --   &  --  &      \\
         & $ \oq \pi     $ & P &  1   &  14  &  4   &  4   &      \\
         & $ \ksq K      $ & P &  --  &   3  &  --  &  --  &      \\
         & $\Gamma $       &  &  103  &  121 &  70  &  112 &      \\
\hline
$2^{+-}$ & $ \omega \pi  $ & D &  .5  &   1  &   1  &   1  &   4  \\
         & $\rho\eta     $ & D &   .1 &  .6  &  .2  &   .1 &  1   \\
         & $\rho\eta^{'} $ & D &   0  &  .02  &   0  &   0  &  0  \\
         & $ K^{\ast} K  $ & D &  .04 &  .2  & .08  & .04  & .6   \\
         & $ \at\pi  $     & P &  .7  &  .9  &  .4  & 130  &      \\
         &                 & F &   0  &  .02 &   0  &  .2  &      \\
         & $ \aoo\pi  $     & P &  3   &  4   &  2   & 45   &      \\
         &                 & F & .01  &  .02 &  0   & .3   &      \\
         & $ \ho\pi  $     & P &  2   &   2  &  1   & 69   &      \\
         &                 & F & .01  & .03  & .01  & .5   &      \\
         & $ \bo\eta  $    & P & .02  &  .5  & .01  & .8   &      \\
         &                 & F &  0   &   0  &   0  &  0   &      \\
         & $ \kt K $       & P & --   & .04  & --   &  --  &      \\
         &                 & F & --   &   0  & --   &  --  &      \\
         & $ \kl K $       & P &  0   &  .03 & 0    & .6   &      \\
         &                 & F &  0   & 0    &  0   &  0   &      \\
         & $ \kh K $       & P & --   & .3   & --   &  --  &      \\
         &                 & F & --   &  0   & --   &  --  &      \\
         & $ \pq \pi  $    & D & .08  &  1   &  .2  & .2   &      \\
         & $ \oq \pi  $    & D & .02  &  .4  &  .04 & .04  &      \\
         & $ \ksq K      $ & D &  --  &  .01 &  --  &  --  &      \\
         & $\Gamma$        &   & 7    &  11 &   5   &  248 &      \\ 
\hline
$0^{-+}$ & $ \rho \pi    $ & P &  37  &  63  &  51  &  47  & 230  \\
         & $ K^{\ast} K  $ & P &   5  &  18  &  10  &   5  &  69  \\
         & $ \rho \omega $ & P &      &      &  \O  &      &      \\
         & $ \ft\pi  $     & D &   1  &   3  & .6   & 8    &      \\
         & $ \fz\pi  $     & S &  62  &  40  & 30   & 62   &      \\
         & $ \at\eta  $    & D &  --  &  .1  & --   & --   &      \\
         & $ \az\eta  $    & S &  --  &  4   & --   & --   &      \\
         & $ \kt K $       & D & --   &   .02& --   &  --  &      \\
         & $ \kz K $       & S & --   & 44   & --   &  --  &      \\
         & $ \rqb \pi $    & P &  3   &  47  & 10   & 10   &      \\
         & $ \ksq K   $    & P &  --  &   5  & --   & --   &      \\
         & $\Gamma $       &  &  108  & 224  & 102  & 132  &      \\
\hline
$1^{+-}$ & $\omega\pi  $   & S &   23 &  19  &  26  &  38  & 118  \\
         &                 & D &   .3 &  .8  &  .4  &  .3  &  2   \\
         & $\rho\eta     $ & S &   15 &  21  &  25  &   22 & 118  \\
         &                 & D &  .07 &  .3  &  .1  &  .06 &  .6  \\
         & $\rho\eta^{'} $ & S &   3  &  8   &  5   &   4  & 25  \\
         &                 & D &  0   &  .01 &  0   &  0   &   0  \\
         & $ K^{\ast} K  $ & S &  27  &  52  & 47   & 36   & 339  \\
         &                 & D &  .02 &  .1  & .04  & .02  & .3   \\
         & $ \at\pi  $     & P & 19   &  26  & 10   & 49   &      \\
         &                 & F &  0   &   .02&  0   &  .1  &      \\
         & $ \aoo\pi  $     & P &  9   &   10 &  5   &   29 &      \\
         & $ \az\pi  $     & P &  3   &   6  &  1   &  26  &      \\
         & $ \ho\pi  $     & P &  \O  &  \O  &  \O  & 95   &      \\
         & $ \bo\eta  $    & P &  \O  &  \O  &  \O  &  1   &      \\
         & $ \kt K $       & P & --   &  1   & --   &  --  &      \\
         &                 & F & --   &  0   & --   &  --  &      \\
         & $ \kl K $       & P &  .04 &  .6  & .02  &  5   &      \\
         & $ \kz K $       & P & --   &  .4  & --   &  --  &      \\
         & $ \kh K $       & P & --   &  .4  & --   &  --  &      \\
         & $ \oq \pi  $    & S &  16  &  82  &  58  & 79   &      \\
         &                 & D & .01  &  .2  &  .02 & .02  &      \\
         & $ \ksq K   $    & S &  --  & 110  &  --  &  --  &      \\
         &                 & D &  --  &  .01 &  --  &  --  &      \\
         & $\Gamma$        &   &  115 & 338  & 177  & 384  &      \\ 
\hline
$0^{+-}$ & $ \aoo\pi  $     & P & \O   &  \O  & \O   & 309  &      \\
         & $ \ho\pi  $     & P & 47   &  45  & 24   &  37  &      \\
         & $ \bo\eta  $    & P & .6   &  12  &  .4  &  .3  &      \\
         & $ \kl K $       & P &  .7  &  10  & .4   &  7   &      \\
         & $ \kh K $       & P & --   &   1  & --   &  --  &      \\
         & $ \pq \pi  $    & S & 60   & 246  &  222 &  312 &      \\
         & $ \kq K    $    & S & --   & 115  & --   &  --  &      \\
         & $\Gamma $       &   & 108  & 429  & 247  &  665 &      \\
\hline
$1^{++}$ & $ \rho \pi    $ & S &  23  &  19  &  26  &  38  &  116 \\
         & $             $ & D &   1  &   3  &   2  &   1  &   8  \\
         & $ K^{\ast} K  $ & S &  14  &  26  &  24  &  18  &  170 \\
         & $             $ & D &  .04 &  .3  &   .09& .04  &   .6 \\
         & $ \rho \omega $ & S &   0  &   0  &   0  &   0  &  47  \\
         &                 & D &   0  &   0  &   0  &   0  &  .03  \\
         & $ \ft\pi  $     & P & 4    &  5   & 2    & 75   &      \\
         &                 & F & .01  & .03  & 0    &  .3  &      \\
         & $ \fo\pi  $     & P &  7   &  9   & 4    &   62 &      \\
         & $ \fz \pi  $      & P &  \O  & \O   & \O   &   4  &      \\
         & $ \bo\pi $      & P &  \O  &  \O  & \O   &      &      \\
         & $ \at\eta  $    & P &  --  &  .9  &  --  &  --  &      \\
         &                 & F &  --  &  0   &  --  &  --  &      \\
         & $ \aoo\eta  $    & P &  .2  &  3   & .09  &   1  &      \\
         & $ \az\eta  $    & P &  --  &  \O  &  --  &  --  &      \\
         & $ \kt K $       & P & --   &  .4  & --   &  --  &      \\
         &                 & F & --   &   0  & --   &  --  &      \\
         & $ \kl K $       & P &  .07 &   1  & .05  &  1   &      \\
         & $ \kz K $       & P & --   &   0  & --   &  --  &      \\
         & $ \kh K $       & P & --   &   .7 & --   &  --  &      \\
         & $ \rqb \pi $    & S & 14   & 80   & 50   & 66   &      \\
         &                 & D & .02  & .6   & .05  & .04  &      \\
         & $ \ksq K   $    & S &  --  & 55   & --   & --   &      \\
         &                 & D &  --  & .01  & --   & --   &      \\
         & $\Gamma$        &  &  63  & 204  & 108  & 269   &      \\

\end{tabular}
\label{table2}
\end{table}

\begin{table}
\caption{Isoscalar Hybrid Decay Modes}
\begin{tabular}{lccccccc}
         &                &   &  alt & high mass & standard & IKP & reduced \\
\hline
$2^{-+}$ & $ K^{\ast} K  $ & P & 1    &   5  &   2  & 1    &  17  \\
         & $ \at \pi $     & S &  52  & 31   &  25  & 45   &      \\
         &                 & D &  .2  & .6   &  .1  & 22   &      \\
         & $ \aoo\pi  $     & D &  .5  &  1   &  .3  & \O   &      \\
         & $ \az\pi  $     & D &  .02 &  .1  & .01  &  .6  &      \\
         & $ \ft\eta  $    & S &  --  &  8   & --   &  --  &      \\
         &                 & D &  --  & .02  & --   &  --  &      \\
         & $ \fo\eta  $    & D &  --  & .02  & --   &  --  &      \\
         & $ \fz\eta  $    & D &  --  &   0  & --   &  --  &      \\
         & $ \kt K $       & S & --   & 11   & --   &  --  &      \\
         &                 & D & --   &  0   & --   &  --  &      \\
         &                 & G & --   &  0   & --   &  --  &      \\
         & $ \kl K $       & D & 0    & .01  &  0   & 0    &      \\
         & $ \kz K $       & D & --   & 0    & --   &  --  &      \\
         & $ \kh K $       & D & --   & 0    & --   &  --  &      \\
         & $ \ksq K   $    & P & --   &  1   &  --  &  --  &      \\
         & $\Gamma$        &   & 54   &  58  &  27  &  69  &      \\
\hline
$1^{-+}$ & $\eta^{'} \eta$ & P &  0   &  0   &  0   &   0  & 10   \\
         & $ K^{\ast} K  $ & P & 1    &  5   &  2   &   1  & 17   \\
         & $ \at\pi  $     & D &  .4  &   1  &  .2  &  \O  &      \\
         & $ \aoo\pi  $     & S &  59  &  30  &  28  &  38  &      \\
         &                 & D &  .3  &  .6  &  .2  &  34  &      \\
         & $ \ft\eta  $    & D &  --  &  .05 & --   &  --  &      \\
         & $ \fo\eta  $    & S &  --  &    8 & --   &  --  &      \\
         &                 & D &  --  &   .01& --   &  --  &      \\
         & $ \kt K $       & D & --   & 0    & --   &  --  &      \\
         & $ \kl K $       & S &  4   & 7    &  2   &   7  &      \\
         &                 & D & 0    & .2   & 0    & 0    &      \\
         & $ \kh K $       & S & --   & 33   & --   &  --  &      \\
         &                 & D & --   &  0   & --   &  --  &      \\
         & $ \pq \pi  $    & P & 8    &  65  &  27  &  27  &      \\
         & $ \euq \eta$    & P & --   &  6   &  --  &  --  &      \\
         & $ \kq K    $    & P & --   &  .8  &  --  &  --  &      \\
         & $ \ksq K   $    & P & --   &   1  &  --  &  --  &      \\
         & $\Gamma $       &   &  73  & 158  &  59  & 107  &      \\
\hline
$0^{-+}$ & $ K^{\ast} K  $ & P & 5    & 18   & 10   &   5  & 69   \\
         & $ \at\pi  $     & D &  2   &  6   &  1   & 16   &      \\
         & $ \az\pi  $     & S & 145  & 114  &  70  & 175  &      \\
         & $ \ft\eta  $    & D &  --  &.2    & --   &  --  &      \\
         & $ \fz\eta  $    & S &  --  & 23   & --   &  --  &      \\
         & $ \kt K $       & D & --   &   .02& --   &  --  &      \\
         & $ \kz K $       & S & --   & 44   & --   &  --  &      \\
         & $ \ksq K   $    & P &      &   5  &      &      &      \\
         & $\Gamma $       &   & 152  & 210  &  81  & 196  &      \\
\hline
$1^{--}$ & $ \rho \pi    $ & P & 28   &  47  & 38   & 35   & 172  \\
         & $ \omega \eta $ & P & 3    &   9  &  6   &  4   &  29  \\
         & $\omega\eta^{'}$& P & .1   &   1  &  .2  & .3   &   .8  \\
         & $ K^{\ast} K  $ & P & 3    &  9   &   5  &  3   &  35  \\
         & $ \bo\pi  $     & S & \O   &  \O  &  \O  &      &      \\
         &                 & D &      &      &  \O  &      &      \\
         & $ \ho\eta  $    & S &      &      &      &  \O  &      \\
         & $ \kt K $       & D & --   & 0    & --   &  --  &      \\
         & $ \kl K $       & S &  6   &  12  &  4   &  11  &      \\
         &                 & D &  0   & .01  & 0    &  0   &      \\
         & $ \kh K $       & S & --   &  17  & --   &  --  &      \\
         &                 & D & --   & 0    & --   &  --  &      \\
         & $ \rqb  \pi  $  & P &  2   &  35  &  8   &  7   &      \\
         & $ \oq \eta $    & P & --   &  .6  &  --  &  --  &      \\
         & $ \ksq K   $    & P & --   &   3  &  --  &  --  &      \\
         & $\Gamma $       &   &  42  & 134  &  61  &  60  &      \\
\hline
$2^{+-}$ & $ \rho \pi    $ & D & 1    &   4  &  2   &  2   &  11  \\
         & $ \omega \eta $ & D & .1   &  .5  &  .2  &   .1 &   1  \\
         & $\omega\eta^{'}$& D &  0   &  .03  &   0  &   0  & 0  \\
         & $ K^{\ast} K  $ & D & .04  &  .2  &  .08 &  .04 & .6   \\
         & $ \bo\pi  $     & P &   4  &  5   & 2    & 164  &      \\
         &                 & F &  .02 &  .07 & .01  &   .8 &      \\
         & $ \ho\eta $     & P & .2   &  .7  & .1   &  6   &      \\
         & $ \kt K $       & P & --   & .04  & --   &  --  &      \\
         &                 & F & --   &   0  & --   &  --  &      \\
         & $ \kl K $       & P &  0   &  .03 & 0    & .6   &      \\
         &                 & F &  0   & 0    &  0   &  0   &      \\
         & $ \kh K $       & P & --   & .3   & --   &  --  &      \\
         &                 & F & --   &  0   & --   &  --  &      \\
         & $ \rqb \pi $    & D &  .02 & .8   &  .06 &  .05 &      \\
         & $ \oq \eta $    & D &  --  &  0   &  --  &  --  &      \\
         & $ \ksq K   $    & D & --   &  .01 &  --  &  --  &      \\
         & $\Gamma$        &   & 5    &  12  &   4  & 166  &      \\
\hline
$1^{+-}$ & $ \rho \pi    $ & S & 70   &  57  &  77  & 114  & 350  \\
         &                 & D & .8   &   2  &   1  &   1  &   6  \\
         & $ \omega \eta $ & S & 15   &  22  &  25  & 22   & 119  \\
         & $             $ & D & .07  &  .3  &  .1  & .06  &  .6  \\
         & $\omega\eta^{'}$& S & 4    &  8   &  5   & 15   & 24   \\
         & $             $ & D & 0    &  .02 &   0  &  0   & 0   \\
         & $ K^{\ast} K  $ & S & 27   &  52  &  47  & 36   & 339  \\
         & $             $ & D &  .02 &   .1 &  .04 & .02  &  .3  \\ 
         & $ \bo\pi  $     & P &  \O  &  \O  &  \O  & 231  &      \\
         & $ \ho\eta  $    & P &  \O  &  \O  & \O   &  9   &      \\
         & $ \kt K $       & P & --   &  1   & --   &  --  &      \\
         &                 & F & --   &  0   & --   &  --  &      \\
         & $ \kl K $       & P &  .04 &  .6  & .02  &  5   &      \\
         & $ \kz K $       & P & --   &  .4  & --   &  --  &      \\
         & $ \kh K $       & P & --   &  .4  & --   &  --  &      \\
         & $ \rqb \pi  $   & S &  42  & 240  &  150 & 199  &      \\
         &                 & D & .01  &  .4  &  .04 & .03  &      \\
         & $ \oq \eta $    & S &  --  & 38   &  --  &  --  &      \\
         &                 & D &  --  &  0   &  --  &  --  &      \\
         & $ \ksq K   $    & S &  --  & 110  &  --  &  --  &      \\
         &                 & D &  --  &  .01 &  --  &  --  &      \\
         & $\Gamma$        &   &  158 & 529  & 305  & 632  &      \\
\hline
$0^{+-}$ & $ \bo\pi  $     & P &  110 &  119 &  56  &  85  &      \\
         & $ \ho \eta $    & P &  4   &  17  &  3   &  2   &      \\
         & $ \kl K $       & P &  .7  &  10  & .4   &  7   &      \\
         & $ \kh K $       & P & --   &   1  & --   &  --  &      \\
         & $ \kq K    $    & S &  --  & 115  &  --  &  --  &      \\
         & $\Gamma$        &   & 115  & 262  &  59  &  94  &      \\
\hline
$1^{++}$ & $ K^{\ast} K  $ & S & 17   &  26  &  24  &  18  & 170  \\
         &                 & D &  .04 &  .3  &  .09 &  .04 &  .6  \\
         & $ \at\pi  $     & P &   10 &  14  &   5  & 179  &      \\
         &                 & F &  .01 & .06  &  .01 &  .4  &      \\
         & $ \aoo\pi  $     & P &   28 & 30   &   14 & 232  &      \\
         & $ \az\pi  $     & P &  \O  &  \O  &   \O &   6  &      \\
         & $ \ft\eta  $    & P &  --  &   1  & --   &  --  &      \\
         &                 & F &  --  &   0  & --   &  --  &      \\
         & $ \fo\eta  $    & P &  --  &   2  & --   &  --  &      \\
         & $ \fz\eta  $    & P & \O   &  \O  & \O   &  --  &      \\
         & $ \kt K $       & P & --   &  .4  & --   &  --  &      \\
         &                 & F & --   &   0  & --   &  --  &      \\
         & $ \kl K $       & P &  .07 &   1  & .05  &  1   &      \\
         & $ \kz K $       & P & --   &   0  & --   &  --  &      \\
         & $ \kh K $       & P & --   &   .7 & --   &  --  &      \\
         & $ \ksq K   $    & S & --   & 55   &  --  &  --  &      \\
         &                 & D &  --  & .01  &  --  &  --  &      \\
         & $\Gamma$        &   &  55  & 130  &  43  & 436  &      \\

\end{tabular}
\label{table3}
\end{table}

\begin{table}
\caption{$s \bar s$ Hybrid Decay Modes}
\begin{tabular}{lccccccc}
         &                &   &  alt & high mass & standard & IKP & reduced \\
\hline
$2^{-+}$  & $K^{\ast} K  $ & P &  6   & 13   &  11  &   8  &  82  \\
         & $ \kt K $       & S &  28  &  29  & 21   & 44   &      \\
         &                 & D &  .03 &  .5  &  .02 &  1   &      \\
         & $ \kl K $       & D &  .2  &  .5  &  .1  &   10 &      \\
         & $ \kz K $       & D & .02  &   .3 &   .01&   .2 &      \\
         & $ \kh K $       & D &  .06 &    .5&  .03 &  .6  &      \\
         & $ \fts\eta $    & S &  --  &  20  &  --  &  --  &      \\ 
         &                 & D &  --  &  .2  &  --  &  --  &      \\
         & $ \fos\eta $    & D &  --  &  .03 &  --  &  --  &      \\ 
         & $ \fzs\eta $    & D &  .01 &   .08&  0   &  .1  &      \\ 
         & $ \ksq K   $    & P &  2   &  27  &  6   &  5   &      \\
         & $\Gamma$        &   &  36  &  91  &  38  &  69  &      \\
\hline
$1^{-+}$  & $ \eta^{'}\eta$& P &  0   &  0   &   0  &   0  &  44  \\
          & $ K^{\ast} K $ & P &  6   & 13   &  11  &   8  &  82  \\
         & $ \kt K $       & D &  .07 & 1    &  .04 &  \O  &      \\
         & $ \kl K $       & S &  14  & 10   &  11  &  14  &      \\
         &                 & D &   3  & 8    &   2  &  21  &      \\
         & $ \kh K $       & D &  83  & 76   & 61   & 121  &      \\
         &                 & D &  .03 &  .2  & .02  &  .4  &      \\
         & $ \fts\eta $    & D &  --  & .04  &  --  &  --  &      \\
         & $ \fos\eta $    & S &  --  & 21   &  --  &  --  &      \\
         &                 & D &  --  & .02  &  --  &  --  &      \\
         & $ \kq K    $    & P &  1   &  45  & 4    &  3   &      \\
         & $ \esq \eta$    & P &  --  &  15  & --   &  --  &      \\
         & $ \ksq K   $    & P &  2   &  27  &  6   &  5   &      \\
         & $\Gamma$        &   & 109  & 216  &  95  & 172  &      \\
\hline
$0^{-+}$ & $ K^{\ast} K  $ & P & 26   & 52   &  46  &  33  & 330  \\
         & $ \kt K $       & D &   .4 &   6  &  .2  &  1   &      \\
         & $ \kz K $       & S &  113 &   117&  83  &  174 &      \\
         & $ \fts\eta $    & D &  --  &  .2  &  --  &  --  &      \\
         & $ \fzs\eta $    & S &  72  &  105 &  64  & 109  &      \\
         & $ \ksq K   $    & P &   7  &  110 &  22  &  18  &      \\
         & $\Gamma$        &   & 218  & 390  & 215  & 335  &      \\
\hline
$1^{--}$ & $ K^{\ast} K  $ & P & 13   & 26   &  23  &  16  & 165  \\
         & $ \phi   \eta $ & P & 2    &  19  &  11  &   3  &  89  \\
         & $ \phi\eta^{'}$ & P & .01  &  2   &  .1  &  .02 &  .5  \\
         & $ \kt K $       & D &  .1  & 2    & .07  &   2  &      \\
         & $ \kl K $       & S &  23  & 16   &  18  &  24  &      \\
         &                 & D &   .2 &  .6  &  .1  &  2   &      \\
         & $ \kh K $       & S &  43  & 40   &  32  &  63  &      \\
         &                 & D &  .1  & .6   &  .04 &  .7  &      \\
         & $ \hos\eta $    & S &      &      & \O   &      &      \\
         &                 & D &      &      & \O   &      &      \\
         &                 & D &  .07 &  .6  & .04  &  .3  &      \\
         & $ \ksq K   $    & P &  3   &   55 &  11  &   9  &      \\
         & $\Gamma$        &   &  84  & 155  &  95  & 120  &      \\
\hline
$2^{+-}$ & $ K^{\ast} K  $ & D &  1   &  3   &   2  &   1  &  13  \\
         & $ \phi \eta   $ & D & .06  &  .8  &  .3  &  .08 &   2  \\
         & $ \phi\eta^{'}$ & D & 0    &  0  &  0   &  0   &   0  \\
         & $ \kt K $       & P &  .3  & 1    &  .2  &  32  &      \\
         &                 & F &   0  & .03  &  0   & .01  &      \\
         & $ \kl K $       & P &  .2  &  .3  &  .1  &  17  &      \\
         &                 & F &  .04 &  .2  &  .02 &  .6  &      \\
         & $ \kh K $       & P &  3   &   8  &   2  &   28 &      \\
         &                 & F &  0   &   0  &  0   &  0   &      \\
         & $ \hos\eta $    & P &  .3  &  2   & .2   &  9   &      \\
         &                 & F &   0  &  0   &  0   &  0   &      \\
         & $ \ksq K   $    & D &  .04 &   2  &  .1  & .08  &      \\
         & $\Gamma$        &   &  5   &  18  &   5  &  79  &      \\
\hline
$1^{+-}$ & $ K^{\ast} K  $ & S & 20   & 19   &  34  &  42  & 247  \\
         &                 & D &  .6  &  2   &   1  &   .6 &   7  \\
         & $ \phi   \eta $ & S &  11  & 63   &  66  &  28  & 523  \\
         &                 & D &  .03 &  .5  &  .2  &  .04 &   1  \\
         & $ \phi \eta^{'}$& S &  2   &  19  &  8  &  3  & 61  \\
         & $             $ & D &  0   &  .02 &   0  &  0   &   0  \\
         & $ \kt K $       & P &  8   & 35   &  5   &  10  &      \\
         &                 & F &  0   & .02  &  0   &  .01 &      \\
         & $ \kl K $       & P &  4   &  5   &  2   &  122 &      \\
         & $ \kz K $       & P &  3   & 14   &  2   &   18 &      \\
         & $ \kh K $       & P &  3   &  8   &  2   &   4  &      \\
         & $ \hos\eta $    & P &  \O  &  \O  &  \O  & 14   &      \\
         & $ \ksq K   $    & S &   39 &  206 &  181 & 201  &      \\
         &                 & D &   .02&    1 &  .06 & .04  &      \\
         & $\Gamma$        &   &  91  & 373  & 301  & 443  &     \\
\hline
$0^{+-}$ & $ \kl K $       & P &  66  & 95   &  43  &  165 &      \\
         & $ \kh K $       & P &  10  &  30  &   6  &  36  &      \\
         & $ \hos \eta$    & P &  8   &  42  &  5   &  4   &      \\
         & $ \kq K    $    & S & 46   & 323  & 205  & 221  &      \\
         & $\Gamma$        &   & 130  & 490  &  259 & 426  &     \\
\hline
$1^{++}$ & $ K^{\ast} K  $ & S & 10   &  9   &  17  &  21  & 123  \\
         &                 & D &  1   &  4   &   2  &   1  &  15  \\
         & $ \kt K $       & P &   3  &  13  &   2  & 27   &      \\
         &                 & F &   0  &  .05 &   0  & .01  &      \\
         & $ \kl K $       & P &   7  &  11  &   5  &  37  &      \\
         & $ \kz K $       & P &   \O &  \O  &   \O &  2   &      \\
         & $ \kh K $       & P &   6  &  16  &   3  &  29  &      \\
         & $ \fts\eta $    & P &  --  &  2   &  --  &  --  &      \\
           &               & F &  --  &  0   &  --  &  --  &      \\
         & $ \fos\eta $    & P &  --  &  4   &  --  &  --  &      \\
         & $ \fzs\eta $    & P &  \O  &  \O  &  \O  &  2   &      \\
         & $ \ksq K   $    & S & 19   & 103  & 90   & 100  &      \\
         &                 & D & .05  & 2    &  .1  & .08  &      \\
         & $\Gamma$        &   &  46  & 164  & 119  & 219  &      \\
\end{tabular}
\label{table4}
\end{table}

\begin{table}
\caption{$c \bar c$ Hybrid Decay Modes}
\begin{tabular}{lccccccc}
         &                &   &  alt & high mass & standard & IKP & reduced \\
\hline
$2^{-+}$ & $ D^{\ast}D $   & P &  .5   &  .1  & .8    &  4   &  19  \\
         & $\dss(2^{+})D$  & S & --   &   9   &  --  &  --  &      \\
         &                 & D & --   &  .2  &  --  &  --  &      \\
         & $\dss(1^{+}_L)D$& D & --   &   .2  &  --  &  --  &      \\
         & $\dss(0^{+})D$  & D & --   &   .2  &  --  &  --  &      \\ 
         & $\dss(1^{+}_H)D$& D & --   &   .2 &  --  &  --  &      \\ 
         & $\Gamma$        &   & .5   &  10    & .8   &  4   &       \\
\hline
$1^{-+}$ & $ D^{\ast}D $   & P &  .5  &  .1  & .8    &  4   &  19  \\
         & $\dss(2^{+})D$  & D & --   &   .5  &  --  &  --  &      \\  
         & $\dss(1^{+}_L)D$& S & --   &   1.2  &  --  &  --  &      \\ 
         &                 & D & --   &   2.5 &  --  &  --  &      \\
         & $\dss(1^{+}_H)D$& S & --   &   25 &  --  &  --  &      \\
         &                 & D & --   &   0  &  --  &  --  &      \\
         & $\Gamma $       &   & .5   &  29  & .8   &  4   &      \\
\hline
$0^{-+}$ & $ D^{\ast}D $   & P &  2   &   .3  & 3   & 16   & 76  \\
         & $\dss(2^{+})D$  & D & --   &   2.5  &  --  &  --  &      \\
         & $\dss(0^{+})D$  & S & --   &   25 &  --  &  --  &      \\
         & $\Gamma $       &   & 2    &  28  & 3    &  16  &      \\
\hline
$1^{--}$ & $ D^{\ast}D $   & P &  1   &  .2  &  1.5    &  8   &  38  \\
         & $\dss(2^{+})D$  & D & --   & 1     &  --  &  --  &      \\
         & $\dss(1^{+}_L)D$& S & --   & 7   &  --  &  --  &      \\
         &                 & D & --   &  .3   &  --  &  --  &      \\
         & $\dss(1^{+}_H)D$& S & --   & 10   &  --  &  --  &      \\
         &                 & D & --   & .2   &  --  &  --  &      \\
         & $\Gamma$        &   & 1    & 19   & 1.5   & 8   &      \\
\hline
$2^{+-}$ & $ D^{\ast}D $   & D & .2   &  .2  &  .3   &  1   &   7 \\
         & $\dss(2^{+})D$  & P & --   &  .5   &  --  &  --  &      \\
         &                 & F & --   & .02  &  --  &  --  &      \\ 
         & $\dss(1^{+}_L)D$& P & --   &  0 &  --  &  --  &      \\
         &                 & F & --   &  0   &  --  &  --  &      \\
         & $\dss(1^{+}_H$)D& P & --   &  3  &  --  &  --  &      \\
         &                 & F & --   &  0   &  --  &  --  &      \\
         & $\Gamma $        &   & .2   & 4   & .3   &  1    &      \\
\hline
$1^{+-}$ & $ D^{\ast}D $   & S &  .3   &   .1 &  .5   &  8   &   12 \\
         &                 & D & .1   &    .1&  .1  &  .5  &  4  \\
         & $\dss(2^{+})D$  & P & --   & 13   &  --  &  --  &      \\
         &                 & F & --   & .01  &  --  &  --  &      \\
         & $\dss(1^{+}_L)D$& P & --   & 2    &  --  &  --  &      \\
         & $\dss(0^{+})D$  & P & --   & 8    &  --  &  --  &      \\
         & $\dss(1^{+}_H)D$& P & --   & 2.5   &  --  &  --  &      \\
        & $\Gamma $        &   & .4   &  26  &  .6   & 8.5  &      \\
\hline
$0^{+-}$ & $\dss(1^{+}_L)D$& P & --   & 25  &  --  &  --  &      \\
         & $\dss(1^{+}_H)D$& P & --   & 15  &  --  &  --  &      \\
         & $\Gamma $       &   & --   & 40  &  --  &  -- &       \\
\hline
$1^{++}$ & $ D^{\ast}D $   & S & .2   &  .1  &  .3   &   1  &  6  \\
         &                 & D & .2   &  .2  &  .3   &   .3  &  8  \\
         & $\dss(2^{+})D$  & P & --   &  5  &  --  &  --  &      \\
         &                 & F & --   & .03   &  --  &  --  &      \\
         & $\dss(1^{+}_L)D$& P & --   & 5   &  --  &  --  &      \\
         & $\dss(0^{+})D$  & P & --   & \O   &  --  &  --  &      \\
         & $\dss(1^{+}_H)D$& P & --   & 5   &  --  &  --  &      \\
         & $\Gamma$        &   &  .4  & 15  & .6   & 1.3  &      \\
\end{tabular}
\label{table5}
\end{table}

\begin{table}
\caption{$b \bar b$  Hybrid Decay Modes}
\begin{tabular}{lccccccc}
         &                &   &  alt & high mass & standard & IKP & reduced \\
\hline
$2^{-+}$ & $ B^{\ast}B $   & P &  .1  &   0  & .5    &  3   & 44  \\
$1^{-+}$ & $ B^{\ast}B $   & P &  .1  &   0  &  .5    & 3     & 44  \\
$0^{-+}$ & $ B^{\ast}B $   & P &  .5   &   0  & 2    & 13   & 177  \\
$1^{--}$ & $ B^{\ast}B $   & P &  .2  &   0  &  1.2   &  7   &  88 \\
$2^{+-}$ & $ B^{\ast}B $   & D &  .08  &  .05  &  .25   & 1   &   22 \\
$1^{+-}$ & $ B^{\ast}B $   & S &  .02 &   .1 &  .2  &  5   &   13 \\
         & $ B^{\ast}B $   & D &  .02  &  .02&  .15  &  .6  &  12  \\
$1^{++}$ & $ B^{\ast}B $   & S &  .01 &  .05  &  .25    &   2  &  7  \\
         & $ B^{\ast}B $   & D &  .1  &  .05  &  .5   &   1  &  24  \\
\end{tabular}
\label{table6}
\end{table}


\section{Discussion}

We proceed to discuss the phenomenology of mainly isovector hybrids made from 
$u,d$ flavored quarks for each $J^{PC}$, as these are expected to be the easiest 
to isolate experimentally.

\subsection{Light hybrids}

\subsubsection{$1^{--} $}

\vskip 0.1in

It was argued in Refs. \cite{page96rad,page97rad} that the $\rho(1450)$  and the
$\omega(1420)/\omega(1600)$ cannot be accommodated within the
phenomenologically successful $^3P_0$ decay model as conventional
mesons -- a hybrid component is needed. This conclusion depends
strongly on the results of the influential data analysis of
Ref. \cite{donnachie94}. The central
problem is that the substantial experimental $a_1\pi$
mode \cite{donnachie94} cannot be accommodated along with other modes of $\rho(1450)$ if
the state is $2\; ^3S_1$ or $^3D_1$ quarkonium. However, if the experimental $a_1\pi$
width of 190 MeV \cite{donnachie94} can be reduced by 50\%, the $\rho(1450)$
can be fitted as $2\; ^3S_1\; q\bar{q}$ \cite{page97rad}.  The IKP model 
predicted that
$a_1\pi$ would be the largest decay mode of a hybrid, consistent with the
data. It is of interest to examine these conclusions here.

For an isovector $1^{--} $ state at 1.5 GeV we calculate for
``standard parameters'' the widths
\begin{tabbing}
XXXXXXXXXX\=XXXXX\=XXXXX\=XXXXX\=XXXXX\=XXXXX\=XXX\kill 
\>$\omega\pi$\>$\rho\eta$\>$K^{\ast}K$\>$a_1\pi$\\
this work\>6\>2\>.6\>15\>MeV\\
IKP model\>5\>1\>.3\>43\>MeV\\
\end{tabbing}
where both models predict $\pi\pi,\;\rho\rho,\; KK,\; h_1\pi$ and
$a_2\pi$ to vanish. For an isoscalar state at 1.5 GeV
\begin{tabbing}
XXXXXXXXXX\=XXXXX\=XXXXX\=XXXXX\=XXXXX\=XXXXX\=XXXXX\=XXXXX\=XXXXX\=XXX\kill 
\>$\rho\pi$\>$\omega\eta$\>$K^{\ast}K$\\
this work\>20\>1\>.6\>MeV\\
IKP model\>17\>1\>.3\>MeV\\
\end{tabbing}
where both models predict $KK$ and $b_1\pi$ to be negligible.

The predictions for the models are very similar, except that the 
$a_1\pi$ mode of the isovector state is smaller in this model.
 However, the ordering of modes according to their relative
sizes remains the same, and $a_1\pi$ remains the dominant channel. It is clear that it becomes
difficult to support the huge experimental $a_1\pi$ mode in both
models.
In the light of this we urge quantification
of this mode at DA$\Phi$NE and JLab (and at a coupled channel analysis currently
in progress at Crystal Barrel\cite{ut}).

If the $\rho(1450)$ has indicated the existence of the
vector hybrid nonet, then we
need to establish which of the other seven multiplets expected nearby should also be visible.
States whose couplings are predicted to
be strong, with highly visible decay channels and moderate widths relative
to the $\rho$ candidate, {\it must} be seen if hybrids are to be established. 
Conversely, channels where 
no signals are seen should be those with signals which are predicted to be weak. 

\vskip 0.2in
\subsubsection{$0^{+-}$}

The clearest signature for a hybrid meson is the appearance
of a flavored state with exotic $J^{PC}$. 
It was noted in the IKP model \cite{page95hybrid} that 
the isovector $0^{+-}$ width is predicted to be large (over 600 MeV
according to Table \ref{table2}). Here the width is 100 -- 250 MeV depending on
parameters, making the state narrower. However, as shown in Table \ref{table2},
if the mass of the state increases, the width may increase dramatically. 
There are accordingly two likely reasons why this state has not yet been observed:
(i) Its mass is higher than 1.8 GeV, making it very wide. This possibility
is underpinned by recent lattice gauge theory calculations
supporting a mass difference of $\sim 0.2\pm 0.2$ GeV between  $0^{+-}$ and
the lowest lying $1^{-+}$ hybrid \cite{lacock96}. (ii) Its decay modes are idiosyncratic. It
can be seen from the table that decays are only to $S + P$ --wave
states, most likely to $\pi(1300)\pi,\; a_1\pi$ and
$h_1\pi$. However, $\pi(1300)$, $a_1$, and $h_1$ are broad states, making
the $0^{+-}$ difficult to isolate.

\subsubsection{$2^{+-}$}

The isovector $2^{+-}$ was predicted to be broad
in the IKP model ($\sim 250$ MeV) \cite{page95hybrid}.
This is especially true if the mass of the state increases, as
indicated by lattice gauge theory calculations,
which suggest a mass difference of $\sim 0.7\pm 0.3$ GeV between  $2^{+-}$ and
$1^{-+}$ levels \cite{lacock96}. However,
in this  model we discover a radically different result: $2^{+-}$ is
$\sim 5$ MeV wide and rises to only $\sim 10$ MeV at 2 GeV. The total width of the $2^{+-}$
hence forms a strong test for the model. Part of the difficulty to
detect the $2^{+-}$ may be that decays to $S+S$ --wave states only
occur in D--wave, and that decay modes like $a_2\pi,\; a_1\pi$ and $h_1\pi$ contain broad
P--wave states. However, in view of the possible narrowness of this
state, we urge experimenters to allow for the exotic $2^{+-}$ wave in
partial wave analyses. Particularly,
$a_2\pi\rightarrow(\rho\pi)\pi\rightarrow 4\pi$ should be studied.

\subsubsection{$1^{-+}$}

An excellent opportunity for isolating exotic hybrids occurs
in the $1^{-+}$ wave.  Recently, there has been several experimental claims 
for $1^{-+}$ signals, most notably by Brookhaven and VES, in two distinct mass regions:
(i) Refs. \cite{lee94,ves96warsaw} sees
a broad structure in the mass region $1.6 - 2.2$ GeV in $f_1\pi$, which is suggestive
of being a composite of two objects at $1.7$ and $2.0$ GeV. It is the
latter that appears to have a resonant phase though they admit that
more data is required for a firm conclusion.
(ii) Ref. \cite{bnl97rhopi} claims a resonance at $1593\pm 8$
MeV with width $168\pm 20$ MeV and Ref. \cite{ves92rhopi} a ``preliminary'' resonance at $1.62\pm
0.02$ GeV with width $0.24\pm 0.05$ GeV. 
We hence study model predictions for $1^{-+}$ states
at 1.6 and 2.0 GeV.

Our expectations for a $J^{PC} = 1^{-+}$ hybrid
at  2.0 GeV are (in MeV)
\begin{tabbing} \label{bnl2}
XXXXXXXXXX\=XXX\=XXXXXXX\=XXXXXX\=XXX\=XXXXXX\=XXX\=XXX\=XXXXXXX\=XXX\kill 
\>$b_1\pi$\>$K_1(1400)K$\>$\eta(1295)\pi$\>$\rho\pi$\>$\rho(1450)\pi$\>$f_1\pi$\>$a_1\eta$\>$K_1(1270)K$\\
this work\>43\>33\>27\>16\>12\>10\>7 \>7 \\
IKP model\>58\>75\>21\>16\>12\>38\>13\>19\\
\end{tabbing}
where we have neglected $K^{\ast} K,\; f_2\pi,\; \pi(1300)\eta,\;
K(1460)K$, and $K^{\ast}(1410)K$ modes which are predicted to be smaller than 5 MeV in both models.
Furthermore, the $\eta\pi,\; \eta^{'}\pi,\; \rho\omega,\; a_2\eta$, and
$K_2^{\ast}(1430)K$ modes are all negligible
in both models. Because of the substantially increased phase space
available relative to a 1.6 GeV hybrid candidate, $P+S$ channels are dominant.
The  model has several modes suppressed relative to the IKP
model. Also note in addition to the important $b_1\pi$ channel, $K_1(1400)K$ emerges as prominent channel,
leading us to suggest the search channel $KK\pi\pi$.

For a $1^{-+}$ state at 1.6 GeV one has
\begin{tabbing}\label{bnl1}
XXXXXXXXXX\=XXXXX\=XXXXX\=XXXXX\=XXXXXXXX\=XXXXX\=XXX\kill 
\>$b_1\pi$\>$\rho\pi$\>$f_1\pi$\>$\eta(1295)\pi$\>$K^{\ast}K$\\
this work\>24\>9\>5 \>2\>.8\>MeV\\
IKP model\>59\>8\>14\>1\>.4\>MeV\\
\end{tabbing}
where both models predict $\eta\pi,\; \eta^{'}\pi,\; \rho\omega$ and $f_2\pi$ to be 0 MeV.
Superficially, the main effect of this model is to make the $P+S$ modes
of a more 
similar size to the $S+S$ modes than they are in the IKP model, 
in agreement with the clear presence of the experimental state
in $\rho\pi$. However, this conclusion is parameter dependent (compare
columns 4 and 6 in Table \ref{table2}). Nevertheless we emphasize the importance of
searching for the hybrid in $\rho\pi$, as well as in the $b_1\pi$ and
$f_1\pi$ channels.
Also, both models concur that $b_1\pi$ should be
primarily focused upon. Such a search has been proposed 
and conditionally approved at JLab  \cite{cebafb1pi}. Although both
models underpredict the total experimental width at $\sim 50-100$ MeV,
we do not consider this significant at the level of accuracy expected
of this model, especially in view of the fact that not all possible
decay modes have been calculated.  

The strong dependence of the partial widths on the hybrid mass is displayed in Fig. 1. 
Note that the ``S+P" selection rule forces this to be true for any hybrids 
in the 2 GeV mass range because decays may only occur to final states
near threshold.

\begin{figure}
\begin{center}
\label{figrho}
\caption{Dominant partial widths of a $1^{-+}$ isovector hybrid at various
hybrid masses. The partial widths to $K_1(1400)K,\; \eta(1295)\pi,\; b_1\pi$
and $\rho\pi$ correspond to the highest to the lowest intersections
with the vertical axis.} 
\leavevmode
\hbox{\epsfxsize=4 in}
\epsfbox{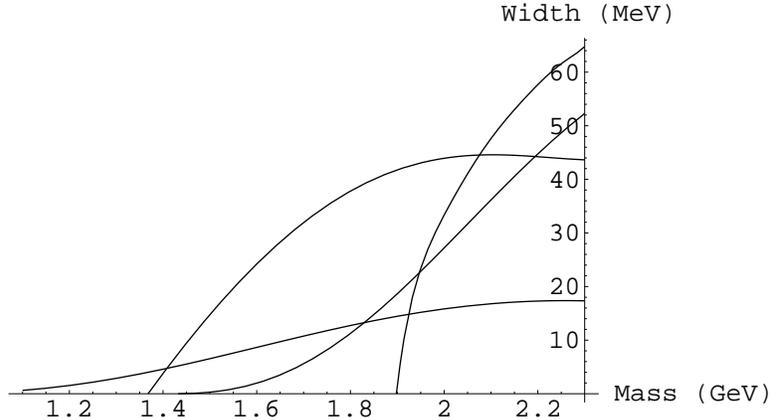}
\end{center}
\end{figure}

It is significant that there is no experimental evidence for
hybrids in the 1.6 -- 2 GeV region in $\eta\pi$ \cite{bnl97etapi} 
which is consistent with the predictions of both models, and is in fact model--independent due to
a relativistic symmetrization selection rule \cite{page97sel1}.
In this context, searches in $\eta\pi$ at JLab and BNL \cite{cebafetapi,bnl97etapi} 
could be disappointing.

More experimental work is needed to clearly establish whether both
$1^{-+}$ signals are solid, and more detailed knowledge of
branching ratios are necessary in order to compare our predictions
with experiment.

\vskip 0.2in

\vskip 0.2in
\subsubsection{$1^{++}$}

\vskip 0.1in

An important model distinction emerges for $1^{++}$ hybrids: we predict widths of 
approximately 
100 MeV, while the IKP model predicts widths larger than
$200$ MeV. We shall argue below that the experimental
evidence for the $a_1(1700)$ indicates that if it is regarded as a
single resonance, then it is not a $1^{++}$ hybrid. Within this model 
{\it either} both the conventional meson and the hybrid are
produced, with the hybrid weaker, {\it or} 
the $1^{++}$ hybrid is higher than 2 GeV in mass, which would push
its width to more than  200 MeV. In either case we expect the dominant decay
channel in this model to be to $1^{++} \rightarrow \rho\pi\rightarrow 3\pi$ or 
$1^{++} \rightarrow \rho(1450)\pi\rightarrow 5 \pi$
(and if phase space allows $K^{\ast}(1410)K$).
Another experimental challenge would be considering
the possibility of two resonances in the $1.6 - 2$ GeV mass region. 

We now argue that the experimental evidence for the $a_1(1700)$ is
consistent with it being a conventional meson. Here we assume for
simplicity that the $a_1(1700)$ is a single resonance, independent of
the channel it is observed in. Current experimental data does not
allow us to go beyond this assumption.  It was noted in 
Ref. \cite{page96rad} that the large D--wave to S--wave ratio for $\rho\pi$ amplitudes
found by VES is consistent with expectations for a $2\; ^3P_1$
conventional meson. It is clear from Table \ref{table2} that the large D--wave
is not explicable for a hybrid in this model or in the IKP model. 
Ref. \cite{page96rad} also predicted a $\rho\pi$ width of $57$ MeV for
$2\; ^3P_1$, while we expect a $\rho\pi$ width of $30$ MeV for a $1.7
$ GeV state. This is consistent with the $2\; ^3P_1$ being strongly produced via the
$\rho\pi$ production vertex sampled at VES. 
This, together with the stronger  
$f_1\pi$ width of the $2\; ^3P_1$ ($18$ MeV), is consistent with the
state observed in $f_1\pi$ \cite{lee94,ves96warsaw} being the $2\; ^3P_1$. VES also 
reported possible
evidence for the $f_0(1370)\pi$ mode. Since the predicted
$f_0(1370)\pi$ width of $2\; ^3P_1$ quarkonium is 2 MeV and that of 
a hybrid is 0
MeV, this supports the weakness of the mode observed.
Recently, VES has reported the observation of a structure in
$\omega\pi^+\pi^-$ at 1.8 GeV that can be identified with the
$a_1(1700)$, 
coupling to the $b_1\pi$ and $\rho\omega$ channels
\cite{ves96warsaw}. In both the current model and the
IKP model, this is inconsistent with the hybrid interpretation, as
the coupling of the hybrid to $b_1\pi$ and $\rho\omega$ is expected to vanish.
In  fact, VES reports an absence of $\rho\omega$ S--wave \cite{zaitpriv}, inconsistent
with the hybrid interpretation where the S--wave dominates the D--wave
(see Table \ref{table2}), but consistent with the $^3 P_0$ model prediction
that the S--wave should be zero (see Eq. (A53) of Ref. \cite{page96rad}). 
Moreover, the $f_1\pi$
channel is dramatically suppressed for the hybrid in this model
in contrast to the IKP model. 
In summary, if we assume that $a_1(1700)$ is a single resonance, it is
consistent with being a conventional meson. Within this assumption, it should be counted 
as one of the successes of this approach that we can explain the
non--observation of the $1^{++}$ hybrid in a way the IKP model cannot. 

\vskip 0.2in
\subsubsection{$0^{-+}$}

\vskip 0.1in

It is clear from Table II that the predictions of this model 
and the IKP model are very similar, except for $f_0(1370)\pi$ which
can vary substantially depending on parameters. 
Refs. \cite{page96rad,page97rad} concluded that the $\pi(1800)$
cannot be understood as a conventional meson in the $^3P_0$ model.
Refs. \cite{page95hybrid,page96panic} concluded that the $\pi(1800)$
can be interpreted as a hybrid meson in the IKP model. 
The current work does not change these conclusions. 
Ref. \cite{ss3} contains a calculation of the widths 
of the $\pi(1800)$ in this model which include below threshold
decays to $K^{\ast}_0(1430)K$ of 85 MeV.\protect\footnote{Some of the 
$K^{\ast}_0(1430)K$ mode predicted
in this model is expected to couple to $f_0(980)\pi$ via
$K^{\ast}_0(1430)K\rightarrow (K\pi)K\rightarrow f_0(980)\pi$
final state interactions, which are known be substantial
experimentally, so that this model estimate is actually less than 85 MeV.} 
It is useful to correlate the decay modes
to experimentally known ratios. Specifically, using the 
VES experimental branching ratios\protect\footnote{The experimentally
measured $KK\pi$ in S--wave is assumed to arise solely from $K^{\ast}_0(1430)K$.} 
\cite{zaitsev95} and correcting
for decays of particles into the specific channels observed by VES
\cite{page97jpsi}, we obtain

\begin{tabbing} \label{bnl2}
XXXXXXXXXX\=XXXXXXXXX\=XXXXXXXX\=XXXXXX\=XXXXX\=XXXXX\=XXXXX\kill 
\>$K^{\ast}_0(1430)K$\>$f_0(1370)\pi$\>$\rho\pi$\>$K^{\ast}K$\>$\rho\omega$\\
Experiment\>$1.0\pm 0.3$ \>$0.9\pm 0.3$\> $<0.36$\> $<0.06$\>$0.4\pm 0.2$\\
this work\>$<0.7$\>0.6\>0.31\>0.05\>0\\
\end{tabbing}
where the model widths evaluated for the ``alternative
parameters'' have been scaled by a common factor to
allow comparison to the experimental ratios deduced in Ref. \cite{page97jpsi}. 
The correspondence is remarkable.

We emphasize that although $\rho\pi$ is suppressed in the data, we
expect the resonance to have a non--negligible coupling to this channel.
The total width is expected to be $\Gamma_{total} \sim 100 - 150$ MeV and is consistent with the
experimental width, since the decay modes $f_0(1500)\pi, \;
f_0(980)\pi$ and $a_0(980)\eta$ that are known to
occur \cite{ves,ves951} have not been computed here and are
experimentally known to give substantial additional
contributions \cite{page97jpsi}. 

One inconsistency with VES data is the $\rho\omega$ mode. It is
significant that the resonance in $\rho\omega$ has a mass $1.732\pm
0.01$ GeV, shifted significantly downward from the usual $\pi(1800)$
mass parameters, and that there are indications of the presence of a 
broad $0^{-+}$ wave \cite{amelin97}. This may signal the presence of $3 ^1 S_0$ light
quark state expected at 1.88 GeV \cite{isgur85} with dominant
decay to $\rho\omega$ \cite{page96rad,page97rad}, removing the
apparent inconsistency with the hybrid interpretation of $\pi(1800)$.

Important tests are now that there should be
a measureable coupling to the $\rho\pi$  channel with only a small
$f_2\pi $ or $K^{\ast}K$ contribution.

\vskip 0.2in

\subsubsection{$2^{-+}$}

\vskip 0.1in

We expect both isovector and isoscalar $2^{-+}$ hybrids to be
narrow, and they should hence be seen. The difference
between the predictions of our approach and the IKP model does not appear
to be substantial, especially when parameters are allowed to vary
(see Table \ref{table2}). The most striking difference between the
models is the isovector $2^{-+}$ decay to $b_1\pi$, which this  model
finds exactly zero.
However, it is fairly small in the IKP model too. 
>From the selection rule forbidding the decay of a spin singlet meson into 
pairs of spin singlets, it follows that the decay of
$^1D_2(Q\bar{Q}) \rightarrow b_1 \pi$ is prevented. Hence the $b_1\pi$ channel may
not be a strong discriminant between hybrid and conventional $2^{-+}$,
as previously suggested \cite{page96rad,page95hybrid}. Recent VES data on the
$2^{-+}$ in $b_1\pi$ does appear to indicate a structure at $1.8$ GeV,
but no firm conclusions are possible at this stage
\protect\cite{ves96warsaw}.
The phenomenology of
the $2^{-+}$ discussed in Refs. \cite{page95hybrid,page96rad} suffices
at this stage: isovector decays to $\rho\pi$ and $f_2\pi$ and
isoscalar decays to $a_2\pi$ remains the dominant signature.

VES noted a $2^{-+}$ structure $\pi_2(2100)$ at $2.09\pm 0.03$ GeV with width
$520\pm 100$ MeV coupling strongly to $f_0(1370)\pi$ but absent in 
$f_2\pi$ and $f_0(980)$ \cite{ves951}, although an
earlier experiment by ACCMOR reported the state in $\rho\pi,\; f_2\pi$ and
$f_0(1370)\pi$ \cite{pdg96}. 
A similar excess may exist in E852 data
\cite{bnl97rhopi}. Theory expects a second radially excited
quarkonium state at 2.13 GeV \cite{isgur85}. 

In the isoscalar sector, evidence exists for a 
$2^{-+}$ resonance at $\sim\; 1.8$ GeV. There are three plausible possibilities for its
interpretation as a conventional quarkonium state:

(i) Light quark $^1D_2$: The light quark $^1D_2$ state
$\eta_2(1645)$ has most likely already been isolated by Crystal Barrel
\cite{cbar1875,cbar1200} and WA102 \cite{wa102}, as interpreted in Ref. \cite{page96rad}.

(ii) $s\bar{s}\; ^1D_2$: This would be a natural assignment for a 
$\sim 1.8$ GeV state, based on the  predicted mass of $1.89$ GeV \cite{isgur85}. However, this assignment appears
troublesome if we consider the fact that it has only been observed in 
final states not\footnote{Although LASS never claimed an isoscalar
$2^{-+}$ resonance, the data appear to indicate an enhancement
at $1.8-1.9$ GeV in the $2^{-+}$ partial wave produced in $K^-p\rightarrow X\Lambda\;
,\;X\rightarrow K^0_S K^{\pm}\pi^{\mp}$ (Fig. 2 of
Ref. \protect\cite{lass88}). Since the production process may enhance
$s\bar{s}$ above light quark production, LASS may have evidence for
the $s\bar{s}$ nature of the enhancement.} containing strangeness. Moreover, there is evidence from 
Crystal Ball and CELLO for an isovector partner at $\sim 1.8$ GeV 
(see the detailed discussion in Ref. \cite{page96rad}), in
contradiction with the $s\bar{s}$ assignment. However, the isovector
partner is not seen in recent analyses from ARGUS \cite{argus}
and L3 \cite{l3}. It is expected that 
E852 would have more to contribute on this subject in the $\rho\pi$
\cite{bnl97rhopi}, $f_1\pi$ and $a_2\eta$ channels \cite{todd}.

(iii) Light quark $2\; ^1D_2$: As observed above, these states are
expected at much higher masses than $\sim 1.8$ GeV, and there is
already evidence for an isovector $2^{-+}$ in the correct mass region.

If future experimental work determines that none of these three
possibilities are viable
interpretations for the  1.8 GeV state, there is a strong possibility that the
$\sim 1.8$ GeV isoscalar state is a hybrid meson. This is because it is
unlikely to be a glueball which is predicted by lattice gauge theory
at $3.0\pm 0.2$ GeV \cite{ukqcd}. We also do not  expect a molecule or
four--quark state in this region, although the state 
may contain a long range $f_2\eta$ component due to its nearness to
the $f_2\eta$ threshold \cite{cbar1875}.

It is hence of interest to determine whether data on the state
is consistent with decays calculated in this work.  
Recently, the WA102 Collaboration reported evidence for two $2^{-+}$ states
in central $pp$ collisions at 450 GeV, which were absent in previous
analyses by WA76 and WA91 at 85, 300 and 450 GeV \cite{wa102}. The 
upper $2^{-+}$ state is found at $1840\pm 25$ MeV with a width of
$200\pm 40$ MeV. The observed decay mode is $a_2\pi$, in accordance
with the predictions of this model and the IKP model.
The Crystal Ball Collaboration reported some time ago a state with undecided
$J^{PC}$ (claimed to be $2^{-+}$) at $1881\pm 32\pm 40$ MeV, with a width of $221\pm 92\pm 44$ MeV, decaying
equally to $a_2\pi$ and $a_0(980)\pi$ \cite{cball}. Similar
conclusions were drawn by the CELLO Collaboration \cite{cball}.

A doubling of isoscalar $2^{-+}$ peaks has also been reported  by 
Crystal Barrel, in the isoscalar sector in
$p\bar p  \to (\eta\pi^{o}\pi^{o})\pi^{o}$ 
\cite{cbar1875}. Masses and widths of $1875\pm 20\pm 35$~MeV and
$200\pm 25\pm 45$~MeV have been reported for the upper $2^{-+}$ state.
The high--mass state $\eta_2(1875)$ has been seen only
in $f_2(1275)\eta$ (only 50 MeV above threshold), and no evidence of it is found in
$a_0 (980)\pi$, $f_0(980)\eta$, or $f_0(1370)\eta$. The absence of the
state in $f_0(1370)\eta$ is consistent with the hybrid interpretation 
(see column 5 of Table \ref{table3}). However, the non--appearance of
the state in $a_2\pi$ appears disasterous at first glance. We would
like to point out here that this is in fact not the
case. Experimentally, 

\begin{equation}
\frac{\Gamma(\eta_2(1875)\rightarrow a_2^0\pi^0 )\; BR(a_2^0\rightarrow
\eta\pi^0)}{\Gamma(\eta_2(1875)\rightarrow f_2\eta)\; BR(f_2\rightarrow \pi^0\pi^0)}
= 0 (+0.8)\;\; \cite{cbar1875}\;\;  \mbox{or}\;\; 0.7\pm 0.4\;\; \cite{cbar1200}
\end{equation}
Employing branching ratios from Ref. \cite{pdg96} and theoretical widths
yields

\begin{equation} 
\frac{\Gamma(\eta_2(1875)\rightarrow a_2\pi )\frac{0.145}{3}}
{\Gamma(\eta_2(1875)\rightarrow f_2\eta)\;\frac{0.847}{3}} \gapproxeq 1.1
(+0.3)
\end{equation}
in both this model and the IKP model for a 1.875 GeV hybrid.
The mean value was obtained for the ``standard parameters'' and the
error corresponds to the ``alternative parameters''\footnote{For a light quark
$^1D_2$ we find a ratio of 1.0
\protect\cite{page96rad} and for a $2\; ^1D_2$ a ratio of 0.7,  
all evaluated for a meson at 1.875 GeV.}. Equality is reached in the
narrow resonance approximation. The ratio appears to be
consistent with the large errors estimated from experiment.

We conclude that although $\eta_2(1875)$ can be $s\bar{s}\; ^1D_2$; it
is equally consistent with the hybrid interpretation. A critical
discriminant between these possibilities would be the experimental
confirmation of an isovector partner \cite{page96rad} since the hybrid
candidate consists of light quarks..




\vskip 0.2in
\subsection{Strangeonium hybrids}

Strangeonium hybrids could be studied by intense photon beams at 
JLab, due to the strong affinity of the photon for $s\bar{s}$.
Vector and
$1^{+-}$ hybrids have non--negligible $\phi\eta$
couplings which could form a good search channel. Moreover,
we note that some non--exotic hybrids are substantially
narrower than their quarkonium partners, e.g. for $J^{PC} = 1^{--}$
the hybrid has a width of $\sim 100$ MeV in both models compared to
the prediction for $^3D_1$ quarkonium of 650 MeV \cite{page96rad}. 
This generates the prospect of photoproduction
of vector states beyond the well known $\phi(1680)$.

When the total widths of all $I=1$, $I=0$ and $s\bar{s}$ hybrids 
listed in Table \ref{table4} are
computed, we find that for ``standard parameters'' the average total
widths of the three flavor varieties are very similar in both models
(although  $I=0$ are about $\sim 30\%$ narrower). This dispells a popular
misconception that $s\bar{s}$ hybrids should be narrower than light
quark hybrids.

\vskip 0.2in
\subsection{Charmonium hybrids}

The widths of charmonium hybrids are suppressed below
$\dss D$ threshold, where only $D^{\ast}D$ and $D^{\ast}_s D_s$
modes are allowed, since these are the only open charm combinations
where the wave functions the two final states are different.
Widths in Table \ref{table5} are in the 1 - 20 MeV range, and hence
surprisingly narrow for charmonia at such high masses. However, when
the hybrids are allowed to become more massive than the $\dss D$ threshold,
the total widths increase drastically (see Figure 2)
to $4 - 40$ MeV for 4.4 GeV hybrids (see column 5 in Table
\ref{table5}). However, in this model (but not in the IKP model
\cite{page95como}) the $2^{+-}$ exotic remains narrow
at $4$ MeV.

\begin{figure}
\begin{center}
\label{figccbar}
\caption{Dominant partial widths of a $1^{-+}\; c\bar{c}$ hybrid at various
masses. The partial widths to $\dss(1^{+}_H)D,\; \dss(1^{+}_L)D,\; 
\dss(2^{+})D$ and $D^{\ast}D$ correspond to the highest to the lowest intersections
with the vertical axis.} 
\leavevmode
\hbox{\epsfxsize=4 in}
\epsfbox{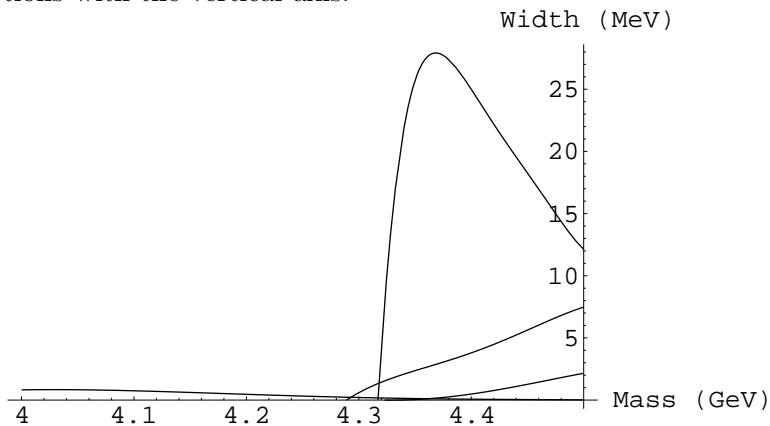}
\end{center}
\end{figure}

\section{Conclusions}

We have explored the implications of the hybrid decay model constructed in
Ref. \cite{ss3}. The model assumes the validity of the flux tube description
of hybrids. The hybrid decay vertex is motivated by the heavy quark limit of the
QCD Hamiltonian. It is essentially given by transverse gluon dissociation into a
$q \bar q$ pair. Thus, the decay model is similar to earlier \cite{Orsay} hybrid
decay models which assumed that constituent gluons  produced $q\bar q$ pairs
in the standard perturbative manner. The main difference is that the hybrid and the
decay mechanism have been written in terms of the degrees of freedom appropriate to
the flux tube model (ie., phonons). In this sense, the model presented here is 
similar to ``$^3S_1$" meson decay models whereas the IKP model is similar to $^3P_0$ models.

This similarity extends to amplitude ratios. Amplitude ratios serve as a sensitive probe
of the decay vertex and may be used to test models. For example, $S/D$ amplitude ratios 
tend to be significantly smaller in $^3P_0$ meson decay models than in $^3S_1$ models due to details of
momentum routing. Because of this it has been shown that $^3P_0$ models are heavily favored by the 
data\cite{geiger94}. A similar situation exists between this model and that of IKP. For example,
the $S/D$ amplitude ratio for $2^{-+}(I=0) \rightarrow a_2 \pi$ is roughly 2 in the IKP model while
it is 250 in this model. Similarly the $S/D$ ratio for $1^{-+}(I=1) \rightarrow b_1\pi$ 
is 5 in the IKP model and 40 in this model. One can envision a time when these ratios 
may be experimentally determined and the models distinguished.

Hybrid states that have small total widths should be accessible
experimentally. We find that for ``standard parameters'' the
total width of the $I=1,\; I=0$ and $s\bar{s}$ $2^{-+}$ hybrids are
less than 100 MeV in both models. Moreover, the same is true for
$I=0$ $1^{--}$ and $s\bar{s}$ $2^{+-}$. The stability of these narrow
widths in both models is significant, and neccesitates experimental
examination of these states. 
There are also states which are less than 100 MeV wide in this model,
but not in the IKP model. These are the $I=1$ and $I=0$ $2^{+-}$, the
$I=0$ and $s\bar{s}$ $1^{-+}$, the $I=0$ $0^{-+}$ and $0^{+-}$. 
In general the IKP model and this one give similar decay widths (in large part because both
obey the spin and S+P selection rules). However they differ dramatically in a few places. The
most obvious is the anomalously narrow width of exotic $2^{+-}$ hybrids predicted by this
model (less than 10 MeV). This surprising result needs to be accounted for in experimental
searches and partial wave analyses. The channel $2^{+-} \rightarrow a_2\pi \rightarrow (\rho\pi) \pi
$ $\rightarrow 4\pi$ is especially important in this regard.

Other differences are in the total widths of the 
$0^{+-}(I=1)$ and $1^{+-}(I=0)$ hybrids, which we predict to be roughly 200 MeV, while IKP
predict values 3 times larger. A larger discrepancy is in the $1^{++}(I=0)$ state which we
predict to be 50 MeV wide, while IKP predict 450 MeV.

Among the conclusions of our survey of interesting hybrid candidates were the 
following.
The $\rho(1450)$ remains enigmatic and further experimental study of this state is vital.
This is especially true of the $a_1 \pi$ mode which appears to be anomalously large.  

Amongst quantum number-exotic hybrids, the isovector $0^{+-}$ appears to be very wide and
thus may be difficult to detect. Alternatively, there is growing evidence for (several) $1^{-+}$
states. We stress the importance of exploring the $b_1\pi$ and $f_1\pi$ channels as well
as $\pi\rho$ and, if the hybrid is heavy enough, $K_1(1400)K$. In fact the latter mode is  
expected to be the largest if the hybrid is heavier than 2.1 GeV.

The $\pi(1800)$ is difficult to accomodate as a conventional meson and makes a likely 
hybrid candidate. Indeed, the experimental branching ratios agree spectacularly with our 
predictions. Alternatively, it appears likely that the $a_1(1700)$ is a $2^3P_1$ 
quarkonium state due to the small S-wave $\pi\rho$ mode and the strong $f_1\pi$ channel.
Finally, we conclude that the $\eta_2(1875)$ can be an $s\bar s$ $^1D_2$ state or a hybrid.
Searching for an isovector partner for this state would therefore be especially interesting.

All $c\bar c$ and $b \bar b$ hybrids are very narrow if they lie within their expected mass
ranges. Since the heavy quarkonium spectrum is well understood, searches for these hybrids
are especially interesting.

In general, all hybrid widths depend strongly on available phase space so that care
should be exercised when employing our results. Furthermore, there can be substantial 
parameter dependence in the predicted widths. The standard and alternative data sets typically
led to predictions differing by 50\% and sometimes as much as 100\%. Finally, the overall
scale is not well known and may change substantially as new information emerges.
We look forward to the day when hybrids and their decays are experimentally well established 
since this is doubtlessly an important step in developing an understanding of the mechanics
of strong QCD and low energy glue.

\acknowledgments

Financial support of the DOE under grants  DE-FG02-96ER40944 (ESS) and 
  DE-FG02-87ER40365 (APS) is acknowledged. 
PRP acknowledges a Lindemann Fellowship from the English Speaking Union to 
visit JLab, where part of this work was completed.
\appendix
\section*{}

The ``standard parameters'' are as follows. All $\beta$'s are those
of Ref. \cite{page95hybrid}, i.e. for $u\bar{u}$, $s\bar{s}$,
$c\bar{c}$, $b\bar{b}$ hybrids 0.27, 0.30, 0.30, 0.34 GeV, for
$\at$, $\aoo$, $\az$, $\bo$, $\ft$, $\fo$, $\fz$, $\ho$, $D^{**}$ 0.34 GeV, for
$\pi(1300)$, $\rho(1450)$, $\omega(1420)$ 0.35 GeV, for $K(1460)$, $K^{\ast}_0(1410)$
0.37 GeV, for $K^{\ast}_2(1430)$, $K_1(1270)$, $ K^{\ast}_0(1430)$, $K_1(1400)$ 0.38 GeV, 
for $\pi$, $\rho$, $\omega$, $D$, $D^{\ast}$  0.39 GeV, for $B$, $B^{\ast}$, $\fts$, $\fos$, $\fzs$
, $\hos$ 0.41 GeV, 
for $\eta_u(1295)$ 0.42 GeV, for $K$, $K^{\ast}$ 0.43 GeV, for $\eta_s(1490)$ 0.45 GeV,
for $\phi(1680)$ 0.46 GeV,  for $\eta$, $\eta^{'}$ 0.47 GeV and for
$\phi$ 0.54 GeV. 
In the case of hybrid decays to
S--wave mesons the widths are zero for $\beta_A=\beta_B$. The width
divided by $(\beta_A^2-\beta_B^2)^2/(\beta_A^2+\beta_B^2)^2$ remains
finite, and is called the ``reduced width''. For hybrid decays to
S--wave mesons we calculate the actual width by multiplying the
reduced width by $(\beta_A^2-\beta_B^2)^2/(\beta_A^2+\beta_B^2)^2$, but
this time we take the $\beta$'s to be those of Ref. \cite{kokoski87}
, i.e. for $\pi$ 0.75 GeV, $\eta,\eta^{'}$ 0.74 GeV, $\rho,\omega$ 0.45
GeV, $\phi$ 0.51 GeV, $K$ 0.71 GeV, $K^{\ast}$ 0.48 GeV, $D$ 0.66 GeV,
$D^{\ast}$ 0.54 GeV, $B$ 0.64 GeV and $B^{\ast}$ 0.57 GeV. We assume
that the quarks that are created may have different mass than
the initial quarks. Specifically, the mass of the $u,s,c,b$ quarks
are assumed to be  $0.33, 0.55, 1.82, 5.12$ GeV. 

We assume $\dss_{0^{++}}$ and 
$\dss_{1^{+H}}$ (high mass $1^+$ state) to have masses of 2.40 and 2.45 GeV 
respectively. The wave functions are taken to be S.H.O.
wave functions except for the hybrid, where a radial prefactor of
$r^{\delta}$, with $\delta = 0.62$ is assumed \cite{page95hybrid}.
The $^{3}P_{1} / \: ^{1}P_{1}$--mixing is $34^{o}$
\protect\cite{isgur85} in the P--wave
kaon sector. $\dss_{1^{+L}}$ / $\dss_{1^{+H}}$ mixing is $41^{o}$. 

The ``alternative parameters'' (also employed in
Ref. \cite{ss3}) change from the preceding as follows.
$\beta$ of all hybrids are $0.3$ GeV. $\beta$ of
$\pi,\rho,\omega,K,K^{\ast},\phi,D,D^{\ast},B,B^{\ast}$
are 0.54, 0.31, 0.31, 0.53, 0.36, 0.43, 0.45, 0.37, 0.43, 0.40 GeV respectively\cite{swanson92}. Other mesons have $\beta = 0.35$ GeV \cite{swanson92}. We allow the final states
to have different $\beta$'s. All other conventions are the same
as for the ``standard parameters''.

Note that the overall normalization of pair creation differs for ``standard''
and ``alternative'' parameters.

\end{document}